\documentclass{midl} 

\usepackage{mwe} 
\usepackage{hyperref}

\jmlryear{2025}
\jmlrvolume{-- 186}
\jmlrworkshop{Full Paper -- MIDL 2025}
\editors{Accepted for publication at MIDL 2025}

\title[PathoSAM]{Segment Anything for Histopathology}

\midlauthor{\Name{Titus Griebel\midljointauthortext{Contributed Equally}\nametag{$^{1,2}$}} \orcid{0009-0008-1752-7788} 
  \Email{titus.griebel@stud.uni-goettingen.de}\\
  \Name{Anwai Archit\midlotherjointauthor\nametag{$^{2}$}} \orcid{0009-0002-9533-8620} \Email{anwai.archit@uni-goettingen.de}\\
  \Name{Constantin Pape\nametag{$^{2,3,4}$}} \orcid{0000-0001-6562-7187} \Email{constantin.pape@informatik.uni-goettingen.de}\\
  \addr $^{1}$ University Medicine Göttingen,  Department of General, Visceral and Pediatric Surgery\\
  \addr $^{2}$ Georg-August-University Göttingen, Institute of Computer Science\\
  \addr $^{3}$ CAIMed - Lower Saxony Center for AI \& Causal Methods in Medicine, Göttingen\\
  \addr $^{4}$ Cluster of Excellence Multiscale Bioimaging (MBExC), Georg-August-University Göttingen\\
}

\begin{document}

\maketitle

\begin{abstract}
Nucleus segmentation is an important analysis task in digital pathology.
However, methods for automatic segmentation often struggle with new data from a different distribution, requiring users to manually annotate nuclei and retrain data-specific models. Vision foundation models (VFMs), such as the Segment Anything Model (SAM), offer a more robust alternative for automatic and interactive segmentation.
Despite their success in natural images, a foundation model for nucleus segmentation in histopathology is still missing. Initial efforts to adapt SAM have shown some success, but did not yet introduce a comprehensive model for diverse segmentation tasks.
To close this gap, we introduce PathoSAM, a VFM for nucleus segmentation, based on training SAM on a diverse dataset.
Our extensive experiments show that it is the new state-of-the-art model for automatic and interactive nucleus instance segmentation in histopathology.
We also demonstrate how it can be adapted for other segmentation tasks, including semantic nucleus segmentation. For this task, we show that it yields results better than popular methods, while not yet beating the state-of-the-art, CellViT.
Our models are open-source and compatible with popular tools for data annotation. We also provide scripts for whole-slide image segmentation.
\end{abstract}


\section{Introduction}
Nucleus segmentation is an important task in computational pathology, with applications in cancer grading, tumor micro-environment analysis and survival prediction \cite{dl-for-histopatho}.
Most approaches rely on supervised learning, typically based on UNets \cite{unet}.
Several methods, e.g. StarDist \cite{stardist-histopatho} or HoVerNet \cite{hovernet}, address domain-specific challenges, such as overlapping nuclei or joint segmentation and classification.
HoVer-UNet \cite{hoverunet}, HoVerNeXt \cite{hovernext}, and InstanSeg \cite{instanseg} focus on efficiency without sacrificing segmentation accuracy.
Despite these developments, existing models struggle to generalize to new conditions, for example due to different staining protocols, imaging devices, or sample extraction quality.
Adapting them to new data involves training new models, which requires domain expertise for precise and exhaustive annotations (drawing each nucleus in new training patches). While tools like QuPath \cite{qupath} are popular for this task, the procedure remains tedious and resource intensive.

Recently, vision foundation models (VFM) for segmentation tasks have been introduced \cite{sam,seem}.
They generalize to different imaging domains and support different segmentation tasks, including automatic and \emph{interactive} segmentation.
In particular, the Segment Anything Model (SAM) \cite{sam} enables interactive segmentation from user input.
SAM has been successfully applied and adapted to biomedical imaging domains \cite{micro-sam, sam-med2d, surgical-sam, medico-sam}, but so far has not been comprehensively adapted to histopathology, where the performance of the original SAM is limited \cite{sam-digital-pathology,sam-path,micro-sam,nasir-sam}.
Prior work used SAM for specific tasks in histopathology.
For example, CellViT \cite{cellvit} used its pretrained encoder for semantic instance segmentation. 
However, it only addresses automatic segmentation, neglecting interactive segmentation, and the available models are trained on a single dataset, PanNuke \cite{pannuke}.
Other approaches, such as \cite{nasir-sam} use SAM for semi-supervised instance segmentation. However, this approach does not leverage the full potential of interactive segmentation with SAM, which enables human-in-the-loop correction and data-specific finetuning \cite{micro-sam}. More recently, BioMedParse \cite{biomedparse}, which is based on SEEM \cite{seem}, has demonstrated text-prompt based semantic segmentation, but lacks support for instance segmentation.
In summary, a VFM for automatic and interactive nucleus segmentation in histopathology is missing.

\begin{figure}[ht!]
\floatconts
  {fig:main}
  {\caption{a) Overview of the PathoSAM: we train generalist models for nucleus segmentation on a combination of 6 datasets. The models support interactive and automatic instance segmentation, are integrated with user-friendly tools, and can be further finetuned for semantic segmentation. b) Automatic instance segmentation results of 12 datasets. PathoSAM (ViT-L) is the best model, despite evaluations favoring other models by choosing the best model per dataset (if multiple versions are available). c) Interactive instance segmentation results of 12 datasets. PathoSAM is the best model, especially for single point prompts.}}
  {\includegraphics[width=0.84\linewidth]{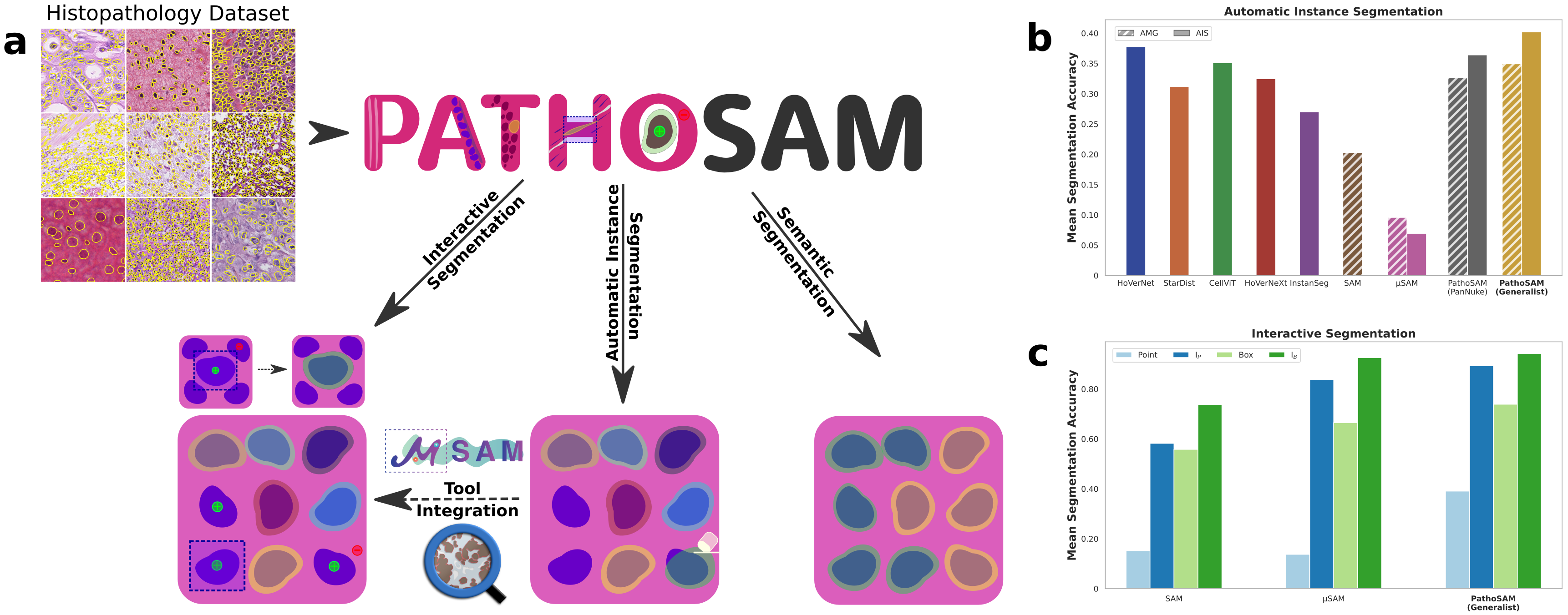}}
  \label{fig:main}
\end{figure}

Here, we present \textit{PathoSAM}, a VFM for nucleus segmentation in histopathology. It supports interactive and automatic segmentation, based on finetuning SAM on a large and diverse annotated dataset.
PathoSAM outperforms other SAM variants for interactive segmentation, is the new state-of-the-art for nucleus instance segmentation, and supports finetuning for semantic segmentation.
PathoSAM is compatible with the user-friendly tools QuPath \cite{qupath} and $\mu$SAM \cite{micro-sam}.
We provide scripts to automatically segment whole-slide-images, together with the rest of our code, at \url{https://github.com/computational-cell-analytics/patho-sam}.
Our contributions and a summary of results are shown in Fig.~\ref{fig:main}.

\section{Methods}
We review interactive (Sec.~\ref{sec:method_sam}) and automatic (Sec.~\ref{sec:method_ais}) instance segmentation with SAM and $\mu$SAM, which form the basis of PathoSAM, our VFM for histopathology (Sec.~\ref{sec:method_vfm}).
We also study automatic semantic segmentation based on PathoSAM (Sec.~\ref{sec:method_semantic}). Our models are trained and evaluated on a large aggregated histopathology dataset (Sec.~\ref{methods:sec_data}).

\subsection{Segment Anything Model and Interactive Segmentation} \label{sec:method_sam}
SAM \cite{sam} has introduced a new formulation for interactive segmentation, where a user can provide input prompts, points (positive or negative), a bounding box or a low-resolution mask, to identify an object.
The model then predicts the corresponding mask by processing the image with the image encoder, a vision transformer (ViT, \cite{vit}), the prompts with the prompt encoder, and the outputs of image encoder and prompt encoder with the mask decoder.
The predictions can be corrected by providing more prompts. See Fig.~\ref{fig:architecture} for an overview of SAM's architecture.

SAM is trained on a large natural image dataset with annotations, using an objective that simulates interactive segmentation. In each iteration, this objective first samples prompts from an annotated mask, predicts the object mask, and then iteratively corrects the prediction with prompts sampled from the annotation.
Predictions and annotations are compared with a loss function for each iteration, and the average loss is used for parameter updates.
We use the training implementation of $\mu$SAM \cite{micro-sam}.

To evaluate interactive segmentation, we automatically derive prompts from annotations to segment the object and then iteratively correct the segmentation, similar to the training objective. We perform 7 correction iterations.
We compute the mean segmentation accuracy (App.~\ref{app:metric}) between annotations and predictions for the initial prompt and corrections. 
We report the results for an initial point prompt, an initial box prompt, and the last correction iteration, when starting from a point ($I_{P}$), and when starting from a box ($I_{B}$).

\subsection{Automatic Segmentation} \label{sec:method_ais}
SAM supports automatic instance segmentation by placing a grid of point prompts across the image, applying the model to each prompt to obtain segmentation masks, and removing overlapping or unlikely masks. This approach is called automatic mask generation (AMG).

$\mu$SAM \cite{micro-sam} introduces an alternative method for automatic instance segmentation, by adding a decoder, similar to UNETR \cite{unetr}, to the model. It predicts foreground probabilities, the distance to the closest object center and the distance to the closest object boundary. During training, the decoder is jointly optimized with the rest of the model, using targets derived from ground-truth annotations.
During inference, the decoder outputs are processed with a seeded watershed to obtain an instance segmentation.
$\mu$SAM has termed this approach automatic instance segmentation (AIS), and has shown that it improves over AMG in terms of segmentation quality and computational efficiency for microscopy data.
We evaluate interactive segmentation (for SAM or variants) with AMG and, if available, AIS, using the mean segmentation accuracy, (App.~\ref{app:metric}). 

\subsection{A vision foundation model for histopathology} \label{sec:method_vfm}
In order to train a VFM model for histopathology, we adopt the training procedure of $\mu$SAM for joint training of interactive (Sec. ~\ref{sec:method_sam}) and automatic (Sec. ~\ref{sec:method_ais}) segmentation. 
We train different models based on the datasets for nucleus segmentation (Sec. ~\ref{methods:sec_data}): three generalist models trained on 6 datasets, one for each of the three image encoders (ViT-B, ViT-L and ViT-H), which we call PathoSAM generalists, and a ViT-B model trained on PanNuke.
The training settings can be found in App.~\ref{app:arch_train}.

\subsection{Semantic Segmentation} \label{sec:method_semantic}
We extend PathoSAM to semantic segmentation, to support the analysis of different cell types.
We add a separate decoder, following the same design as in Sec.~\ref{sec:method_ais}.
See also Fig.~\ref{fig:architecture} for the combined architecture.
This decoder predicts one output channel per semantic class, including background, and is optimized based on the cross entropy loss.
We train and evaluate it on PanNuke (\cite{pannuke}, Sec.~\ref{methods:sec_data}), which provides semantic annotations for 5 classes.
Here, we compare 4 different training strategies, which start from a PathoSAM generalist: (i) freeze the image encoder, initialize the new decoder from scratch; (ii) finetune the image encoder initialize the new decoder from scratch; (iii) freeze the image encoder, initialize the new decoder with the AIS decoder weights; (iv) finetune the image encoder, initialize the new decoder with the AIS decoder weights.
Semantic segmentation is evaluated with the class-frequency weighted dice score,
see App.~\ref{app:metric}.

\subsection{Data} \label{methods:sec_data}
We assemble 14 publicly available histopathology datasets to train and evaluate PathoSAM.
We use 6 datasets \cite{cpm,lizard,monuseg,pannuke,puma,tnbc} with tissue images stained with hematoxylin and eosin (H\&E) staining and nucleus instance annotations to train our generalists. For each of these datasets we reserve a separate test split. An additional model is trained on PanNuke \cite{pannuke}, which is the most popular dataset for nucleus segmentation. This dataset is also used for semantic segmentation, since it provides semantic annotations for 5 classes.
We use the remaining 8 datasets \cite{hovernet,cryonuseg,glas,lynsec,monusac,nuclick,nuinsseg,srsa-net} for out-of-domain evaluation, i.e to test our models on data which was not directly represented in the training set.
Among these, two datasets contain different segmentation tasks, lymphocyte segmentation \cite{srsa-net} and gland segmentation \cite{glas}.
They are used for separate evaluation experiments.
Details on the datasets can be found in App.~\ref{app:data}.

\begin{figure}[ht]
\floatconts
  {fig:res_automatic}
  {\caption{a) Automatic instance segmentation results for 8 datasets. Italic font indicates that the corresponding training split of the dataset was used for our generalist training, while bold font indicates that it was not. b) Interactive segmentation results for 8 datasets (same as in a)) for PathoSAM and other SAM variants. We report the interactive segmentation quality for initial point and box prompts as well as correction for 7 iterations after an initial point ($I_{P}$) and box ($I_{B}$).}}
  {\includegraphics[width=0.8\textwidth]{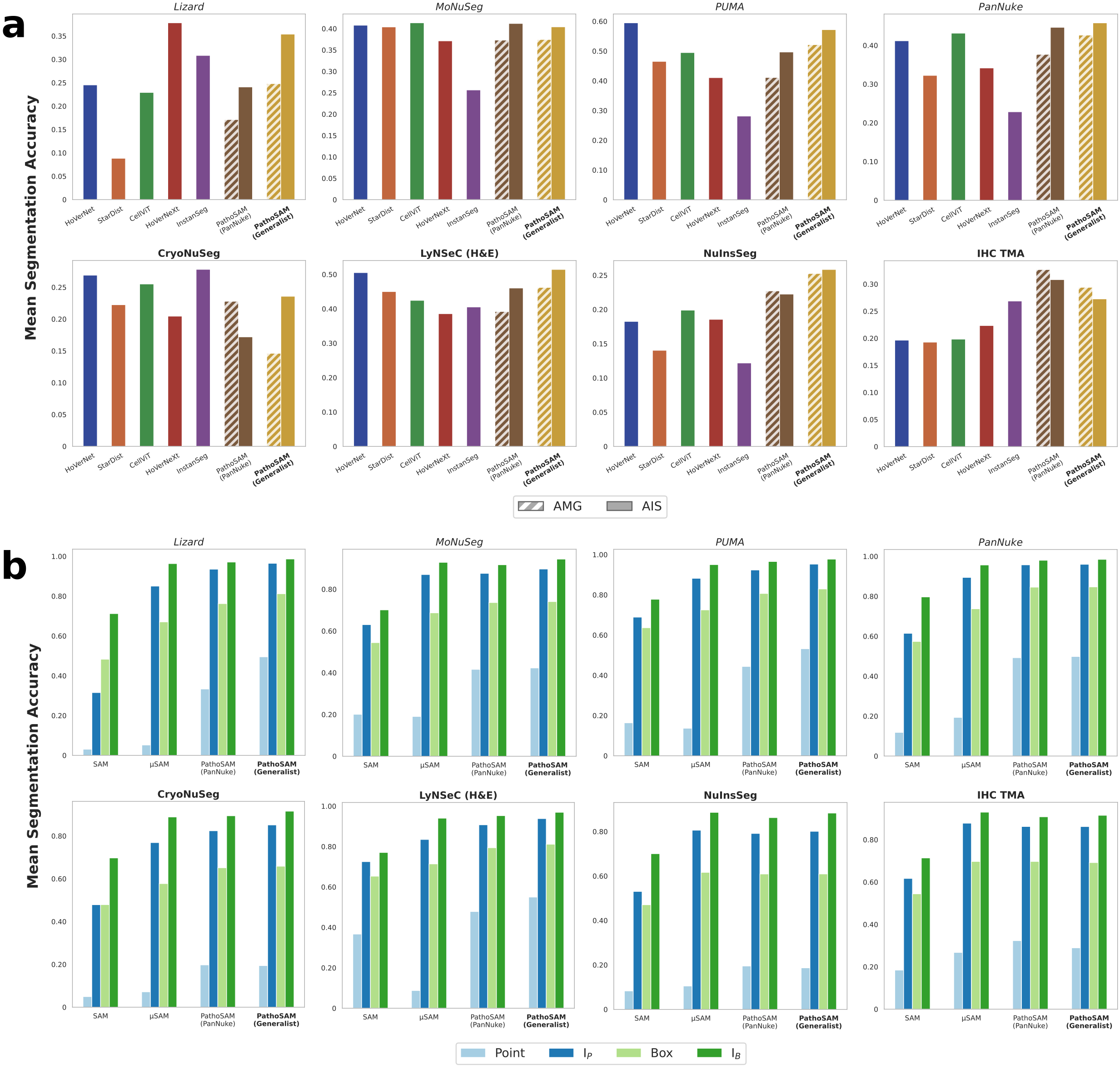}}
  \label{fig:res_automatic}
\end{figure}

\section{Results}
We compare PathoSAM to state-of-the-art methods for automatic instance segmentation (Sec.~\ref{sec:res_automatic}), interactive  segmentation (Sec.~\ref{sec:res_interactive}), and for semantic segmentation (Sec.~\ref{sec:res_downstream}).

\subsection{Automatic Instance Segmentation} \label{sec:res_automatic}

We compare PathoSAM (generalist) with 5 different methods for automatic instance segmentation, HoVerNet \cite{hovernet}, StarDist \cite{stardist-histopatho}, CellViT \cite{cellvit}, HoVerNeXt \cite{hovernext}, InstanSeg \cite{instanseg}, see App. \ref{app:autoseg} details on the other methods.
We evaluate these methods on 12 different datasets, 6 that were part of the training set for our generalists, using separate test splits for evaluation, and 6 that were not.
The methods HoverNet, HoverNext and CellViT offer different model versions. For them, we evaluate all available versions and report the results for the best version per dataset.
Fig.~\ref{fig:res_automatic} a) shows the results on 8 selected datasets, for the ViT-L PathoSAM generalist. Fig.~\ref{fig:main} b) shows the average over all datasets. The detailed results are given in App. \ref{app:autoseg}.

We observe that PathoSAM with AIS is the best overall model, \emph{despite} our reporting favoring methods with multiple versions.
On the ''in-domain`` datasets, PathoSAM is either the best or on par with the best method.
For the ''out-of-domain`` datasets it shows remarkable generalization, outperforming other methods for NuInsSeg (H\&E-stained images) and IHC-TMA (IHC-stained tissue microarrays). CryoNuSeg (cryo-sectioned H\&E-stained images) is the only dataset where it underperforms, see Fig.~\ref{fig:cryonuseg-samples} for details.
We include the comparison to PathoSAM (PanNuke), to investigate the advantage of training on multiple datasets. The effectiveness of this approach is validated, as the generalist model is on par or better for all datasets except IHC-TMA.
We also see that AIS is superior to AMG in most cases. Hence this approach is clearly preferable, also due to its better efficiency \cite{micro-sam}.
The results in \ref{fig:autoseg-quanti} show that differences between ViT-B, ViT-L and ViT-H are small, with ViT-L reaching the highest score.
Furthermore, we investigate the segmentation quality across overlap thresholds in Fig.~\ref{fig:autoseg-f1-precision-recall} and the influence of post-processing parameters on the precision-recall trade-off in Fig.~\ref{fig:heatmap-lynsec-he}.
We also study the segmentation results for neutrophils, which have complex nuclear shapes, in Fig.~\ref{fig:neutrophils-seg}.

\subsection{Interactive Instance Segmentation} \label{sec:res_interactive}
We compare PathoSAM (generalist, ViT-B), with other SAM variants for interactive segmentation.
Specifically, we compare it with the original SAM \cite{sam} and with $\mu$SAM \cite{micro-sam}, which was finetuned on a large microscopy dataset. We use the same 12 datasets as before.
The results for 8 selected datasets are reported in Fig.~\ref{fig:res_automatic} b) and the overall results in Fig.~\ref{fig:main} c).
Detailed results are given in App. \ref{app:intseg}.
We use the evaluation procedure introduced by $\mu$SAM, see also Sec.~\ref{sec:method_sam},
reporting the mean segmentation accuracy for segmentation with a point prompt, a box prompt, iterative correction after an initial point, and iterative correction after an initial box.

PathoSAM outperforms the other models when using a single point or box prompt.
The segmentation quality after iterative correction of PathoSAM and $\mu$SAM are on par; they are clearly better than SAM.
Notably, their segmentation quality after correction is almost perfect.
The strong performance of $\mu$SAM demonstrates that SAM variants can generalize beyond their training set for interactive segmentation.
This is also apparent for CryoNuSeg, where PathoSAM improves drastically through iterative correction. 

\subsection{Other Segmentation Tasks} \label{sec:res_downstream}
\begin{figure}[ht]
\floatconts
  {fig:downstream_res}
  {\caption{Segmentation of lymphocytes a) and glands b) with different SAM variants.
  We report measures for automatic and interactive segmentation, the right hand side shows qualitative examples from the dataset with automatic segmentation results. In both cases, PathoSAM (specialist) is trained on a separate split, using default SAM for weight initialization. c) Semantic segmentation results on PanNuke, comparing PathoSAM with other models. The right hand side shows qualitative examples from all methods.}}
  {\includegraphics[width=0.85\textwidth]{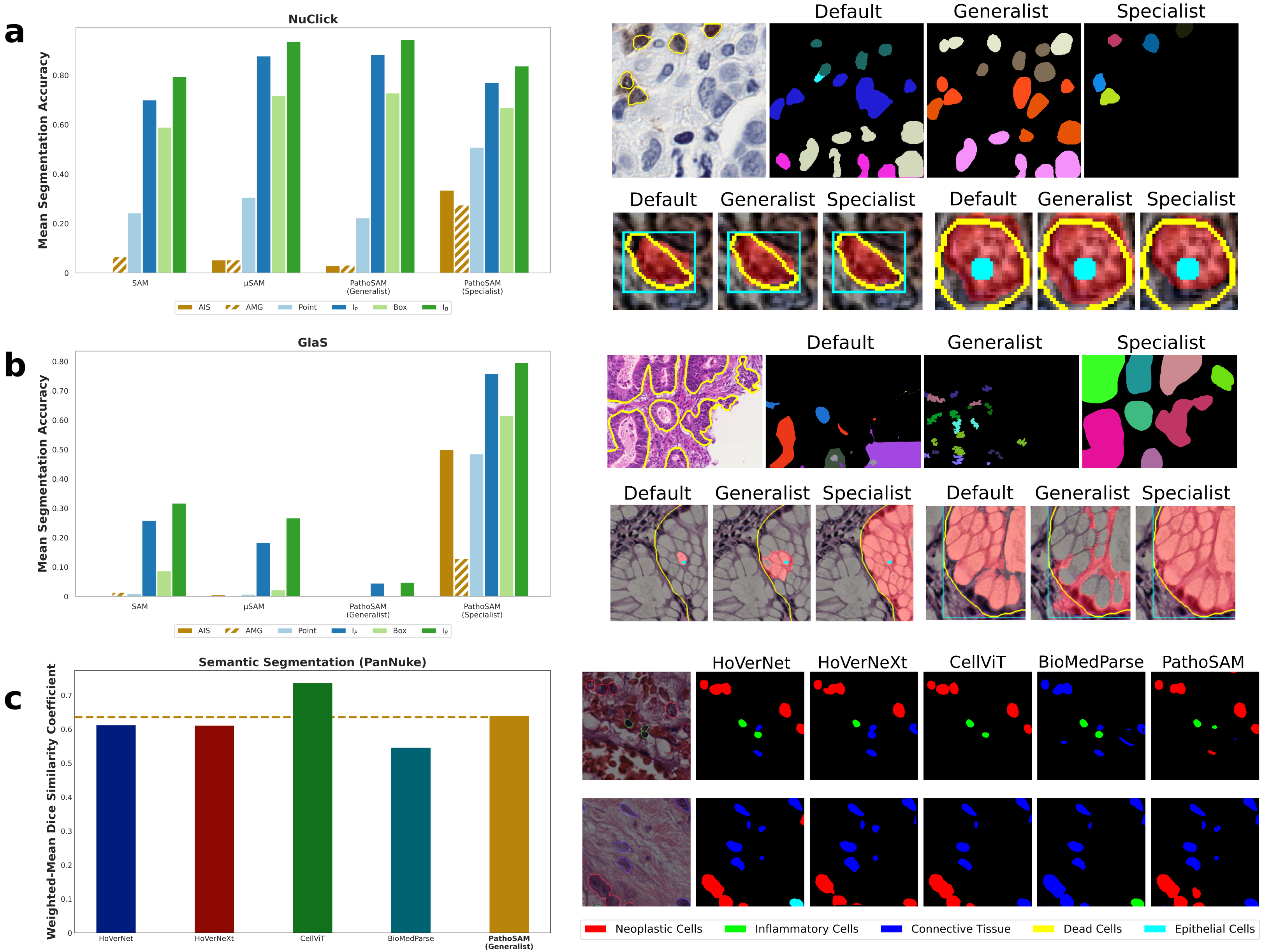}}
  \label{fig:downstream_res}
\end{figure}

We also evaluate PathoSAM on three other segmentation tasks.
First, we evaluate segmentation of different structures, lymphocytes in IHC-stained tissue images \cite{nuclick} and glands in H\&E-stained colon tissue images \cite{glas}.
Here, we compare the automatic and interactive segmentation with SAM, $\mu$SAM, and PathoSAM. We also finetune a new model, PathoSAM (specialist), on the training splits of the respective datasets.
Quantitative and qualitative results are shown in Fig.~\ref{fig:downstream_res} a) and b).
For lymphocytes, SAM, $\mu$SAM and PathoSAM (generalist) cannot automatically segment the data well, but provide good interactive segmentation.
Conversely, the specialist model yields better automatic segmentation, but is worse for interactive segmentation.
The reason for the difference in automatic segmentation is clear: the lymphocytes only make up a fraction of the nuclei in the images.
In automatic segmentation, the non-specific models segment all nuclei, resulting in a bad score. 
The specialist model distinguishes lymphocytes from other nuclei, thus providing a better result.
It is unclear why it performs worse in interactive segmentation.
Overall, this task may be better formulated as a semantic instance segmentation problem, see also next paragraph, to avoid penalizing the segmentation of all nuclei.
For gland segmentation, the specialist outperforms all other SAM variants.
The glands have a different appearance and size compared to nuclei, see the images on the right, explaining this observation.
In both cases, the specialist was trained based on the default SAM, which provided the best results among the non-specific models.
Another architecture based on SAM has been proposed for gland segmentation \cite{zhang2024glandsam}.
However, this approach addresses semantic segmentation, not instance segmentation, so we do not compare to it here.
Overall, these experiments show that neither of the available SAM variants are sufficient to address general instance segmentation tasks beyond nuclei. Finetuning for a specific task using our recipe can however be used to obtain a good tailored model.

We also evaluate PathoSAM for semantic nucleus segmentation on PanNuke \cite{pannuke}. For this application, we train an additional decoder based on the PathoSAM generalist, see Sec.~\ref{sec:method_semantic} for details.
The comparison to other methods trained for this task on PanNuke are shown in Fig.~\ref{fig:downstream_res} c).
Here, we observe that PathoSAM is the second best model, after CellViT, the clear state-of-the-art. HoverNet and HoverNext are on par and better than BioMedParse.
Additional results for semantic segmentation are given in \ref{fig:semantic-seg} and App. \ref{app:autoseg}, reporting the individual class scores and, for PathoSAM, the result for different training strategies. Here, the main observation is that only CellViT correctly segments the ``dead cells'' minority class (\textless 2 \% of the nuclei). Furthermore, finetuning the encoder and training the new decoder from scratch yields the best results for PathoSAM. 

\subsection{Practical Use of PathoSAM} \label{sec:practical}

We evaluate the practical utility of PathoSAM by conducting a qualitative user study for interactive data annotation. 
Here, we check integration with two user-friendly tools: QuPath \cite{qupath}, using the SAM-API plugin \cite{sugawara2023training}, and $\mu$SAM \cite{micro-sam}.
We use the tools to annotate nuclei in \cite{jano-user-study}, which contains nuclei in H\&E stained breast tissue images.
The qualitative results, comparing SAM and PathoSAM, are shown in Fig.~\ref{fig:user_study_qupath} and \ref{fig:user_study_micro_sam}.
We see clear improvements with PathoSAM in both tools: it captures nuclei better from single point or box prompts.

\begin{figure}[hb]
\floatconts
  {fig:wsi_seg}
  {\caption{Segmentation on WSI. The tile shape is 512 $\times$ 512, with an overlap of 64 pixels. a) shows an example for automatic instance segmentation on data from \url{https://openslide.cs.cmu.edu/download/openslide-testdata/} (32,914 $\times$ 46,000 pixels). b) gives resource demand and runtime for the different inference steps using GPU (tested with an A100 with 80GB VRAM) and CPU. We indicate the resources available for the CPU; whereas the entire GPU VRAM available is utilized using batch size of 30. The minimal GPU VRAM requirement is 5GB.}}
  {\includegraphics[width=\textwidth]{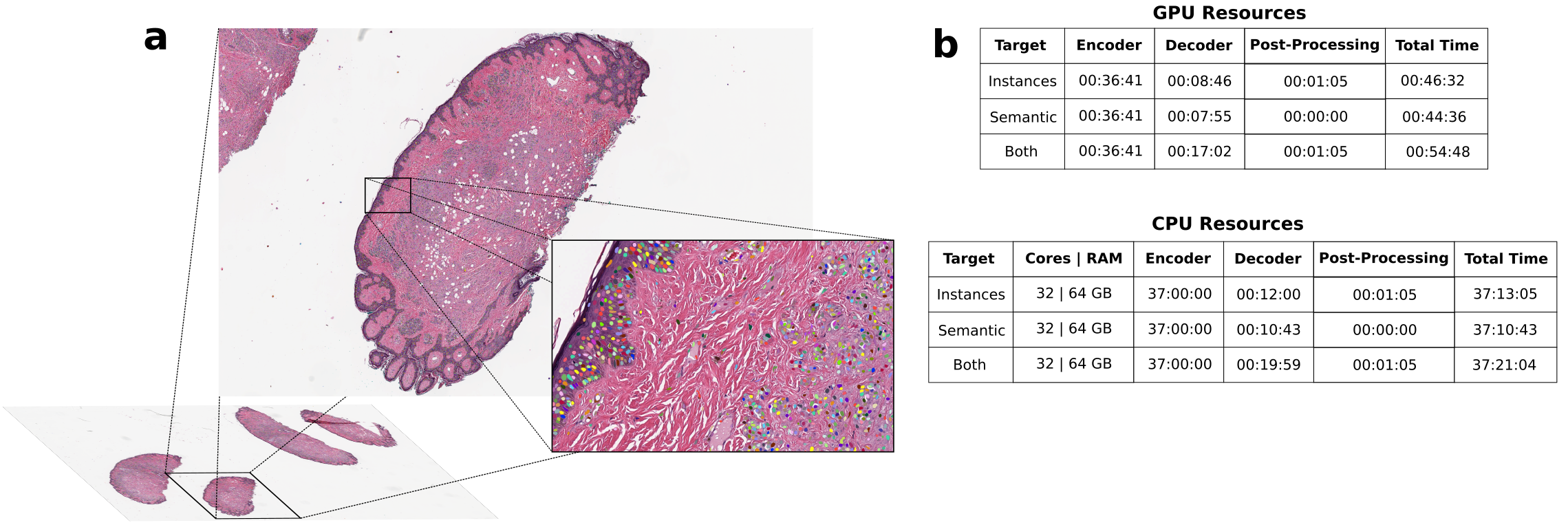}}
  \label{fig:wsi_seg}
\end{figure}

We also implement tile-and-stitch based prediction to apply PathoSAM to whole-slide images (WSI) and enable automatic segmentation in practice.
By default, we use a tile shape of 512 $\times$ 512 pixels, which corresponds to the training patch shape of PathoSAM, with an overlap of 64 pixels on each side.
See Fig.~\ref{fig:wsi_seg} for an example and for an overview of the resources needed for WSI inference.
Segmentation of a WSI runs in less than 1 hour on a GPU, requiring ca. 5 GB of VRAM per tile. It takes much longer on a CPU (ca. 37 hrs).
Our implementation on the GPU has been optimized. The encoder and decoder are applied with a batch size of 1 on CPU resource, otherwise it is automatically selected depending on available VRAM. Increasing the batch size to process multiple tiles in parallel speeds up inference, especially on GPUs with large VRAM.

\section{Discussion}
Our model, PathoSAM, is the new state-of-the-art for interactive and automatic instance segmentation of nuclei in histopathology.
To our knowledge, it is the first foundation model for this task, as it generalizes to diverse settings with a single model.
We believe that it will have great practical impact by speeding up many nucleus analysis tasks. To support this use, we provide integration with user-friendly tools and scripts for processing WSIs.

\paragraph{Limitations:} PathoSAM does not improve segmentation for other tasks in histopathology, e.g. gland or lymphocyte segmentation (see Sec. ~\ref{sec:res_downstream}).
It is thus not yet a general purpose foundation model for histopathology segmentation, but only for nucleus segmentation. However, we show that our finetuning methodology can be used to obtain improved models for other tasks as well; our tool integration provides fast data annotation for such cases.
For semantic segmentation we found that PathoSAM yields good results, but that it is not yet competitive with CellViT, the state-of-the-art.
We believe that this is due to the sophisticated procedure CellViT uses to oversample patches containing minority class annotations.
In contrast, we train without any over-sampling or loss balancing.
Notably, we have only evaluated this task on a single dataset, so it is unclear how any of the methods  for semantic segmentation generalize.

\paragraph{Future Work:} We plan to improve PathoSAM for semantic segmentation by (i) implementing a joint training procedure for instance and semantic segmentation, (ii) over-sampling minority classes (see prev. paragraph), and (iii) assembling a large dataset for semantic training, including annotations from \cite{puma} and \cite{hovernet}. This would provide a versatile foundation model for both instance and semantic segmentation. Furthermore, we will explore training foundation models based on SAM2 \cite{sam2}. Ultimately, we would like to provide a foundation model that addresses further segmentation tasks, such as gland or tissue level segmentation, within a single model. This would however require a multi-scale segmentation approach to exploit tissue-level features that are more prominent after ``zooming out''.

\clearpage  
\midlacknowledgments{
The work of Titus Griebel was supported by the Göttingen Promotionskolleg für Medizinstudierende, funded by Else Kröner-Fresenius-Stiftung.
The work of Anwai Archit was funded by the Deutsche Forschungsgemeinschaft (DFG, German Research Foundation) - PA 4341/2-1. 
Constantin Pape is supported by the German Research Foundation (Deutsche Forschungsgemeinschaft, DFG) under Germany’s Excellence Strategy - EXC 2067/1-390729940.
This work is supported by the Ministry of Science and Culture of Lower Saxony through funds from the program zukunft.niedersachsen of the Volkswagen Foundation for the 'CAIMed – Lower Saxony Center for Artificial Intelligence and Causal Methods in Medicine' project (grant no. ZN4257).
It was also supported by the Google Research Scholarship “Vision Foundation Models for Bioimage Segmentation”.
We gratefully acknowledge the computing time granted by the Resource Allocation Board and provided on the supercomputer Emmy at NHR@G{\"o}ttingen as part of the NHR infrastructure, under the project nim00007.
We would like to thank Sebastian von Haaren for suggestions
on improving data visualizations.
}

\bibliography{midl25_186}

\begin{thebibliography}{40}
\providecommand{\natexlab}[1]{#1}
\providecommand{\url}[1]{\texttt{#1}}
\expandafter\ifx\csname urlstyle\endcsname\relax
  \providecommand{\doi}[1]{doi: #1}\else
  \providecommand{\doi}{doi: \begingroup \urlstyle{rm}\Url}\fi

\bibitem[Alemi~Koohbanani et~al.(2020)Alemi~Koohbanani, Jahanifar, Zamani~Tajadin, and Rajpoot]{nuclick}
Navid Alemi~Koohbanani, Mostafa Jahanifar, Neda Zamani~Tajadin, and Nasir Rajpoot.
\newblock Nuclick: A deep learning framework for interactive segmentation of microscopic images.
\newblock \emph{Medical Image Analysis}, 65:\penalty0 101771, October 2020.
\newblock ISSN 1361-8415.
\newblock \doi{10.1016/j.media.2020.101771}.
\newblock URL \url{http://doi.org/10.1016/j.media.2020.101771}.

\bibitem[Archit et~al.(2025{\natexlab{a}})Archit, Freckmann, Nair, Khalid, Hilt, Rajashekar, Freitag, Teuber, Buckley, von Haaren, Gupta, Dengel, Ahmed, and Pape]{micro-sam}
Anwai Archit, Luca Freckmann, Sushmita Nair, Nabeel Khalid, Paul Hilt, Vikas Rajashekar, Marei Freitag, Carolin Teuber, Genevieve Buckley, Sebastian von Haaren, Sagnik Gupta, Andreas Dengel, Sheraz Ahmed, and Constantin Pape.
\newblock Segment anything for microscopy.
\newblock \emph{Nature Methods}, 22\penalty0 (3):\penalty0 579–591, February 2025{\natexlab{a}}.
\newblock ISSN 1548-7105.
\newblock \doi{10.1038/s41592-024-02580-4}.
\newblock URL \url{http://doi.org/10.1038/s41592-024-02580-4}.

\bibitem[Archit et~al.(2025{\natexlab{b}})Archit, Freckmann, and Pape]{medico-sam}
Anwai Archit, Luca Freckmann, and Constantin Pape.
\newblock Medicosam: Towards foundation models for medical image segmentation, 2025{\natexlab{b}}.
\newblock URL \url{https://arxiv.org/abs/2501.11734}.

\bibitem[Bankhead et~al.(2017)Bankhead, Loughrey, Fernández, Dombrowski, McArt, Dunne, McQuaid, Gray, Murray, Coleman, James, Salto-Tellez, and Hamilton]{qupath}
Peter Bankhead, Maurice~B. Loughrey, José~A. Fernández, Yvonne Dombrowski, Darragh~G. McArt, Philip~D. Dunne, Stephen McQuaid, Ronan~T. Gray, Liam~J. Murray, Helen~G. Coleman, Jacqueline~A. James, Manuel Salto-Tellez, and Peter~W. Hamilton.
\newblock Qupath: Open source software for digital pathology image analysis.
\newblock \emph{Scientific Reports}, 7\penalty0 (1), December 2017.
\newblock ISSN 2045-2322.
\newblock \doi{10.1038/s41598-017-17204-5}.
\newblock URL \url{http://doi.org/10.1038/s41598-017-17204-5}.

\bibitem[Baumann et~al.(2024)Baumann, Dislich, Rumberger, Nagtegaal, Martinez, and Zlobec]{hovernext}
Elias Baumann, Bastian Dislich, Josef~Lorenz Rumberger, Iris~D. Nagtegaal, Maria~Rodriguez Martinez, and Inti Zlobec.
\newblock Hover-next: A fast nuclei segmentation and classification pipeline for next generation histopathology.
\newblock In \emph{Medical Imaging with Deep Learning}, 2024.
\newblock URL \url{https://openreview.net/forum?id=3vmB43oqIO}.

\bibitem[Caicedo et~al.(2019)Caicedo, Goodman, Karhohs, Cimini, Ackerman, Haghighi, Heng, Becker, Doan, McQuin, Rohban, Singh, and Carpenter]{dsb}
Juan~C. Caicedo, Allen Goodman, Kyle~W. Karhohs, Beth~A. Cimini, Jeanelle Ackerman, Marzieh Haghighi, CherKeng Heng, Tim Becker, Minh Doan, Claire McQuin, Mohammad Rohban, Shantanu Singh, and Anne~E. Carpenter.
\newblock Nucleus segmentation across imaging experiments: the 2018 data science bowl.
\newblock \emph{Nature Methods}, 16\penalty0 (12):\penalty0 1247–1253, October 2019.
\newblock ISSN 1548-7105.
\newblock \doi{10.1038/s41592-019-0612-7}.
\newblock URL \url{http://doi.org/10.1038/s41592-019-0612-7}.

\bibitem[Cheng et~al.(2023)Cheng, Ye, Deng, Chen, Li, Wang, Su, Huang, Chen, Jiang, Sun, He, Zhang, Zhu, and Qiao]{sam-med2d}
Junlong Cheng, Jin Ye, Zhongying Deng, Jianpin Chen, Tianbin Li, Haoyu Wang, Yanzhou Su, Ziyan Huang, Jilong Chen, Lei Jiang, Hui Sun, Junjun He, Shaoting Zhang, Min Zhu, and Yu~Qiao.
\newblock Sam-med2d, 2023.
\newblock URL \url{https://arxiv.org/abs/2308.16184}.

\bibitem[Deng et~al.(2023)Deng, Cui, Liu, Yao, Remedios, Bao, Landman, Tang, Wheless, Coburn, Wilson, Wang, Fogo, Yang, and Huo]{sam-digital-pathology}
Ruining Deng, Can Cui, Quan Liu, Tianyuan Yao, Lucas~Walker Remedios, Shunxing Bao, Bennett~A. Landman, Yucheng Tang, Lee~E. Wheless, Lori~A. Coburn, Keith~T. Wilson, Yaohong Wang, Agnes~B. Fogo, Haichun Yang, and Yuankai Huo.
\newblock Segment anything model ({SAM}) for digital pathology: Assess zero-shot segmentation on whole slide imaging.
\newblock In \emph{Medical Imaging with Deep Learning, short paper track}, 2023.
\newblock URL \url{https://openreview.net/forum?id=lUZGyTRzxq}.

\bibitem[Dosovitskiy et~al.(2021)Dosovitskiy, Beyer, Kolesnikov, Weissenborn, Zhai, Unterthiner, Dehghani, Minderer, Heigold, Gelly, Uszkoreit, and Houlsby]{vit}
Alexey Dosovitskiy, Lucas Beyer, Alexander Kolesnikov, Dirk Weissenborn, Xiaohua Zhai, Thomas Unterthiner, Mostafa Dehghani, Matthias Minderer, Georg Heigold, Sylvain Gelly, Jakob Uszkoreit, and Neil Houlsby.
\newblock An image is worth 16x16 words: Transformers for image recognition at scale.
\newblock In \emph{International Conference on Learning Representations}, 2021.
\newblock URL \url{https://openreview.net/forum?id=YicbFdNTTy}.

\bibitem[Gamper et~al.(2019)Gamper, Koohbanani, Benes, Khuram, and Rajpoot]{pannuke}
Jevgenij Gamper, Navid~Alemi Koohbanani, Ksenija Benes, Ali Khuram, and Nasir Rajpoot.
\newblock Pannuke: an open pan-cancer histology dataset for nuclei instance segmentation and classification.
\newblock In \emph{European Congress on Digital Pathology}, pages 11--19. Springer, 2019.
\newblock URL \url{https://doi.org/10.1007/978-3-030-23937-4_2}.

\bibitem[Goldsborough et~al.(2024)Goldsborough, Philps, O'Callaghan, Inglis, Leplat, Filby, Bilen, and Bankhead]{instanseg}
Thibaut Goldsborough, Ben Philps, Alan O'Callaghan, Fiona Inglis, Leo Leplat, Andrew Filby, Hakan Bilen, and Peter Bankhead.
\newblock Instanseg: an embedding-based instance segmentation algorithm optimized for accurate, efficient and portable cell segmentation, 2024.
\newblock URL \url{https://arxiv.org/abs/2408.15954}.

\bibitem[Graham et~al.(2019)Graham, Vu, Raza, Azam, Tsang, Kwak, and Rajpoot]{hovernet}
Simon Graham, Quoc~Dang Vu, Shan E~Ahmed Raza, Ayesha Azam, Yee~Wah Tsang, Jin~Tae Kwak, and Nasir Rajpoot.
\newblock Hover-net: Simultaneous segmentation and classification of nuclei in multi-tissue histology images.
\newblock \emph{Medical Image Analysis}, 58:\penalty0 101563, December 2019.
\newblock ISSN 1361-8415.
\newblock \doi{10.1016/j.media.2019.101563}.
\newblock URL \url{http://doi.org/10.1016/j.media.2019.101563}.

\bibitem[Graham et~al.(2021)Graham, Jahanifar, Azam, Nimir, Tsang, Dodd, Hero, Sahota, Tank, Benes, Wahab, Minhas, Raza, El~Daly, Gopalakrishnan, Snead, and Rajpoot]{lizard}
Simon Graham, Mostafa Jahanifar, Ayesha Azam, Mohammed Nimir, Yee-Wah Tsang, Katherine Dodd, Emily Hero, Harvir Sahota, Atisha Tank, Ksenija Benes, Noorul Wahab, Fayyaz Minhas, Shan E.~Ahmed Raza, Hesham El~Daly, Kishore Gopalakrishnan, David Snead, and Nasir~M. Rajpoot.
\newblock Lizard: A large-scale dataset for colonic nuclear instance segmentation and classification.
\newblock In \emph{Proceedings of the IEEE/CVF International Conference on Computer Vision (ICCV) Workshops}, pages 684--693, October 2021.
\newblock URL \url{https://openaccess.thecvf.com/content/ICCV2021W/CDPath/html/Graham_Lizard_A_Large-Scale_Dataset_for_Colonic_Nuclear_Instance_Segmentation_and_ICCVW_2021_paper.html}.

\bibitem[Hatamizadeh et~al.(2022)Hatamizadeh, Tang, Nath, Yang, Myronenko, Landman, Roth, and Xu]{unetr}
Ali Hatamizadeh, Yucheng Tang, Vishwesh Nath, Dong Yang, Andriy Myronenko, Bennett Landman, Holger~R. Roth, and Daguang Xu.
\newblock Unetr: Transformers for 3d medical image segmentation.
\newblock In \emph{Proceedings of the IEEE/CVF Winter Conference on Applications of Computer Vision (WACV)}, pages 574--584, January 2022.
\newblock URL \url{https://openaccess.thecvf.com/content/WACV2022/papers/Hatamizadeh_UNETR_Transformers_for_3D_Medical_Image_Segmentation_WACV_2022_paper}.

\bibitem[H\"{o}rst et~al.(2024)H\"{o}rst, Rempe, Heine, Seibold, Keyl, Baldini, Ugurel, Siveke, Gr\"{u}nwald, Egger, and Kleesiek]{cellvit}
Fabian H\"{o}rst, Moritz Rempe, Lukas Heine, Constantin Seibold, Julius Keyl, Giulia Baldini, Selma Ugurel, Jens Siveke, Barbara Gr\"{u}nwald, Jan Egger, and Jens Kleesiek.
\newblock Cellvit: Vision transformers for precise cell segmentation and classification.
\newblock \emph{Medical Image Analysis}, 94:\penalty0 103143, May 2024.
\newblock ISSN 1361-8415.
\newblock \doi{10.1016/j.media.2024.103143}.
\newblock URL \url{http://doi.org/10.1016/j.media.2024.103143}.

\bibitem[Janowczyk and Madabhushi(2016)]{jano-user-study}
Andrew Janowczyk and Anant Madabhushi.
\newblock Deep learning for digital pathology image analysis: A comprehensive tutorial with selected use cases.
\newblock \emph{Journal of Pathology Informatics}, 7\penalty0 (1):\penalty0 29, January 2016.
\newblock ISSN 2153-3539.
\newblock \doi{10.4103/2153-3539.186902}.
\newblock URL \url{http://doi.org/10.4103/2153-3539.186902}.

\bibitem[Kirillov et~al.(2023)Kirillov, Mintun, Ravi, Mao, Rolland, Gustafson, Xiao, Whitehead, Berg, Lo, Dollar, and Girshick]{sam}
Alexander Kirillov, Eric Mintun, Nikhila Ravi, Hanzi Mao, Chloe Rolland, Laura Gustafson, Tete Xiao, Spencer Whitehead, Alexander~C. Berg, Wan-Yen Lo, Piotr Dollar, and Ross Girshick.
\newblock Segment anything.
\newblock In \emph{Proceedings of the IEEE/CVF International Conference on Computer Vision (ICCV)}, pages 4015--4026, October 2023.
\newblock URL \url{https://openaccess.thecvf.com/content/ICCV2023/html/Kirillov_Segment_Anything_ICCV_2023_paper.html}.

\bibitem[Kumar et~al.(2017)Kumar, Verma, Sharma, Bhargava, Vahadane, and Sethi]{monuseg}
Neeraj Kumar, Ruchika Verma, Sanuj Sharma, Surabhi Bhargava, Abhishek Vahadane, and Amit Sethi.
\newblock A dataset and a technique for generalized nuclear segmentation for computational pathology.
\newblock \emph{IEEE Transactions on Medical Imaging}, 36\penalty0 (7):\penalty0 1550–1560, July 2017.
\newblock ISSN 1558-254X.
\newblock \doi{10.1109/tmi.2017.2677499}.
\newblock URL \url{http://doi.org/10.1109/tmi.2017.2677499}.

\bibitem[Mahbod et~al.(2021)Mahbod, Schaefer, Bancher, L\"{o}w, Dorffner, Ecker, and Ellinger]{cryonuseg}
Amirreza Mahbod, Gerald Schaefer, Benjamin Bancher, Christine L\"{o}w, Georg Dorffner, Rupert Ecker, and Isabella Ellinger.
\newblock Cryonuseg: A dataset for nuclei instance segmentation of cryosectioned h\&e-stained histological images.
\newblock \emph{Computers in Biology and Medicine}, 132:\penalty0 104349, May 2021.
\newblock ISSN 0010-4825.
\newblock \doi{10.1016/j.compbiomed.2021.104349}.
\newblock URL \url{http://doi.org/10.1016/j.compbiomed.2021.104349}.

\bibitem[Mahbod et~al.(2024)Mahbod, Polak, Feldmann, Khan, Gelles, Dorffner, Woitek, Hatamikia, and Ellinger]{nuinsseg}
Amirreza Mahbod, Christine Polak, Katharina Feldmann, Rumsha Khan, Katharina Gelles, Georg Dorffner, Ramona Woitek, Sepideh Hatamikia, and Isabella Ellinger.
\newblock Nuinsseg: A fully annotated dataset for nuclei instance segmentation in h\&e-stained histological images.
\newblock \emph{Scientific Data}, 11\penalty0 (1), March 2024.
\newblock ISSN 2052-4463.
\newblock \doi{10.1038/s41597-024-03117-2}.
\newblock URL \url{http://doi.org/10.1038/s41597-024-03117-2}.

\bibitem[Naji et~al.(2024)Naji, Sancere, Simon, B\"{u}ttner, Eich, Lohneis, and Bożek]{lynsec}
Hussein Naji, Lucas Sancere, Adrian Simon, Reinhard B\"{u}ttner, Marie-Lisa Eich, Philipp Lohneis, and Katarzyna Bożek.
\newblock Holy-net: Segmentation of histological images of diffuse large b-cell lymphoma.
\newblock \emph{Computers in Biology and Medicine}, 170:\penalty0 107978, March 2024.
\newblock ISSN 0010-4825.
\newblock \doi{10.1016/j.compbiomed.2024.107978}.
\newblock URL \url{http://doi.org/10.1016/j.compbiomed.2024.107978}.

\bibitem[Naylor et~al.(2019)Naylor, Laé, Reyal, and Walter]{tnbc}
Peter Naylor, Marick Laé, Fabien Reyal, and Thomas Walter.
\newblock Segmentation of nuclei in histopathology images by deep regression of the distance map.
\newblock \emph{IEEE Transactions on Medical Imaging}, 38\penalty0 (2):\penalty0 448–459, February 2019.
\newblock ISSN 1558-254X.
\newblock \doi{10.1109/tmi.2018.2865709}.
\newblock URL \url{http://doi.org/10.1109/tmi.2018.2865709}.

\bibitem[Ravi et~al.(2025)Ravi, Gabeur, Hu, Hu, Ryali, Ma, Khedr, R{\"a}dle, Rolland, Gustafson, Mintun, Pan, Alwala, Carion, Wu, Girshick, Dollar, and Feichtenhofer]{sam2}
Nikhila Ravi, Valentin Gabeur, Yuan-Ting Hu, Ronghang Hu, Chaitanya Ryali, Tengyu Ma, Haitham Khedr, Roman R{\"a}dle, Chloe Rolland, Laura Gustafson, Eric Mintun, Junting Pan, Kalyan~Vasudev Alwala, Nicolas Carion, Chao-Yuan Wu, Ross Girshick, Piotr Dollar, and Christoph Feichtenhofer.
\newblock {SAM} 2: Segment anything in images and videos.
\newblock In \emph{The Thirteenth International Conference on Learning Representations}, 2025.
\newblock URL \url{https://openreview.net/forum?id=Ha6RTeWMd0}.

\bibitem[Ronneberger et~al.(2015)Ronneberger, Fischer, and Brox]{unet}
Olaf Ronneberger, Philipp Fischer, and Thomas Brox.
\newblock \emph{U-Net: Convolutional Networks for Biomedical Image Segmentation}, page 234–241.
\newblock Springer International Publishing, 2015.
\newblock ISBN 9783319245744.
\newblock \doi{10.1007/978-3-319-24574-4_28}.
\newblock URL \url{http://doi.org/10.1007/978-3-319-24574-4_28}.

\bibitem[Schuiveling et~al.(2025)Schuiveling, Liu, Eek, Breimer, Suijkerbuijk, Blokx, and Veta]{puma}
Mark Schuiveling, Hong Liu, Daniel Eek, Gerben~E Breimer, Karijn P~M Suijkerbuijk, Willeke A~M Blokx, and Mitko Veta.
\newblock A novel dataset for nuclei and tissue segmentation in melanoma with baseline nuclei segmentation and tissue segmentation benchmarks.
\newblock \emph{GigaScience}, 14, 2025.
\newblock ISSN 2047-217X.
\newblock \doi{10.1093/gigascience/giaf011}.
\newblock URL \url{http://doi.org/10.1093/gigascience/giaf011}.

\bibitem[Sirinukunwattana et~al.(2017)Sirinukunwattana, Pluim, Chen, Qi, Heng, Guo, Wang, Matuszewski, Bruni, Sanchez, B\"{o}hm, Ronneberger, Cheikh, Racoceanu, Kainz, Pfeiffer, Urschler, Snead, and Rajpoot]{glas}
Korsuk Sirinukunwattana, Josien~P.W. Pluim, Hao Chen, Xiaojuan Qi, Pheng-Ann Heng, Yun~Bo Guo, Li~Yang Wang, Bogdan~J. Matuszewski, Elia Bruni, Urko Sanchez, Anton B\"{o}hm, Olaf Ronneberger, Bassem~Ben Cheikh, Daniel Racoceanu, Philipp Kainz, Michael Pfeiffer, Martin Urschler, David~R.J. Snead, and Nasir~M. Rajpoot.
\newblock Gland segmentation in colon histology images: The glas challenge contest.
\newblock \emph{Medical Image Analysis}, 35:\penalty0 489–502, January 2017.
\newblock ISSN 1361-8415.
\newblock \doi{10.1016/j.media.2016.08.008}.
\newblock URL \url{http://doi.org/10.1016/j.media.2016.08.008}.

\bibitem[Sofroniew et~al.(2024)Sofroniew, Lambert, Bokota, Nunez-Iglesias, Sobolewski, Sweet, Gaifas, Evans, Burt, Doncila~Pop, Yamauchi, Weber~Mendon\c{c}a, Buckley, Vierdag, Royer, Can~Solak, Harrington, Ahlers, Althviz~Moré, Amsalem, Anderson, Annex, Boone, Bragantini, Bussonnier, Caporal, Eglinger, Eisenbarth, Freeman, Gohlke, Gunalan, Har-Gil, Harfouche, Hilsenstein, Hutchings, Lauer, Lichtner, Liu, Liu, Lowe, Marconato, Martin, McGovern, Migas, Miller, Muñoz, M\"{u}ller, Nauroth-Kreß, Palecek, Pape, Perlman, Pevey, Peña-Castellanos, Pierré, Pinto, Rodríguez-Guerra, Ross, Russell, Ryan, Selzer, Smith, Smith, Sofiiuk, Soltwedel, Stansby, Vanaret, Wadhwa, Weigert, Windhager, Winston, and Zhao]{napari}
Nicholas Sofroniew, Talley Lambert, Grzegorz Bokota, Juan Nunez-Iglesias, Peter Sobolewski, Andrew Sweet, Lorenzo Gaifas, Kira Evans, Alister Burt, Draga Doncila~Pop, Kevin Yamauchi, Melissa Weber~Mendon\c{c}a, Genevieve Buckley, Wouter-Michiel Vierdag, Loic Royer, Ahmet Can~Solak, Kyle I.~S. Harrington, Jannis Ahlers, Daniel Althviz~Moré, Oren Amsalem, Ashley Anderson, Andrew Annex, Peter Boone, Jordão Bragantini, Matthias Bussonnier, Clément Caporal, Jan Eglinger, Andreas Eisenbarth, Jeremy Freeman, Christoph Gohlke, Kabilar Gunalan, Hagai Har-Gil, Mark Harfouche, Volker Hilsenstein, Katherine Hutchings, Jessy Lauer, Gregor Lichtner, Ziyang Liu, Lucy Liu, Alan Lowe, Luca Marconato, Sean Martin, Abigail McGovern, Lukasz Migas, Nadalyn Miller, Hector Muñoz, Jan-Hendrik M\"{u}ller, Christopher Nauroth-Kreß, David Palecek, Constantin Pape, Eric Perlman, Kim Pevey, Gonzalo Peña-Castellanos, Andrea Pierré, David Pinto, Jaime Rodríguez-Guerra, David Ross, Craig~T. Russell, James Ryan, Gabriel Selzer,
  MB~Smith, Paul Smith, Konstantin Sofiiuk, Johannes Soltwedel, David Stansby, Jules Vanaret, Pam Wadhwa, Martin Weigert, Jonas Windhager, Philip Winston, and Rubin Zhao.
\newblock napari: a multi-dimensional image viewer for python, 2024.
\newblock URL \url{https://zenodo.org/doi/10.5281/zenodo.14427406}.

\bibitem[Sugawara(2023)]{sugawara2023training}
Ko~Sugawara.
\newblock Training deep learning models for cell image segmentation with sparse annotations.
\newblock \emph{BioRxiv}, pages 2023--06, 2023.
\newblock URL \url{https://doi.org/10.1101/2023.06.13.544786}.

\bibitem[Tommasino et~al.(2024)Tommasino, Russo, Rinaldi, and Ciompi]{hoverunet}
Cristian Tommasino, Cristiano Russo, Antonio~Maria Rinaldi, and Francesco Ciompi.
\newblock “hover-unet”: Accelerating hovernet with unet-based multi-class nuclei segmentation via knowledge distillation.
\newblock In \emph{2024 IEEE International Symposium on Biomedical Imaging (ISBI)}, page 1–4. IEEE, May 2024.
\newblock \doi{10.1109/isbi56570.2024.10635755}.
\newblock URL \url{http://doi.org/10.1109/ISBI56570.2024.10635755}.

\bibitem[van~der Laak et~al.(2021)van~der Laak, Litjens, and Ciompi]{dl-for-histopatho}
Jeroen van~der Laak, Geert Litjens, and Francesco Ciompi.
\newblock Deep learning in histopathology: the path to the clinic.
\newblock \emph{Nature Medicine}, 27\penalty0 (5):\penalty0 775–784, May 2021.
\newblock ISSN 1546-170X.
\newblock \doi{10.1038/s41591-021-01343-4}.
\newblock URL \url{http://doi.org/10.1038/s41591-021-01343-4}.

\bibitem[Verma et~al.(2021)Verma, Kumar, Patil, Kurian, Rane, Graham, Vu, Zwager, Raza, Rajpoot, Wu, Chen, Huang, Wang, Jung, Brown, Liu, Liu, Jahromi, Khani, Montahaei, Baghshah, Behroozi, Semkin, Rassadin, Dutande, Lodaya, Baid, Baheti, Talbar, Mahbod, Ecker, Ellinger, Luo, Dong, Xu, Yao, Lv, Feng, Xu, Zunair, Hamza, Smiley, Yin, Fang, Srivastava, Mahapatra, Trnavska, Zhang, Narayanan, Law, Yuan, Tejomay, Mitkari, Koka, Ramachandra, Kini, and Sethi]{monusac}
Ruchika Verma, Neeraj Kumar, Abhijeet Patil, Nikhil~Cherian Kurian, Swapnil Rane, Simon Graham, Quoc~Dang Vu, Mieke Zwager, Shan E.~Ahmed Raza, Nasir Rajpoot, Xiyi Wu, Huai Chen, Yijie Huang, Lisheng Wang, Hyun Jung, G.~Thomas Brown, Yanling Liu, Shuolin Liu, Seyed Alireza~Fatemi Jahromi, Ali~Asghar Khani, Ehsan Montahaei, Mahdieh~Soleymani Baghshah, Hamid Behroozi, Pavel Semkin, Alexandr Rassadin, Prasad Dutande, Romil Lodaya, Ujjwal Baid, Bhakti Baheti, Sanjay Talbar, Amirreza Mahbod, Rupert Ecker, Isabella Ellinger, Zhipeng Luo, Bin Dong, Zhengyu Xu, Yuehan Yao, Shuai Lv, Ming Feng, Kele Xu, Hasib Zunair, Abdessamad~Ben Hamza, Steven Smiley, Tang-Kai Yin, Qi-Rui Fang, Shikhar Srivastava, Dwarikanath Mahapatra, Lubomira Trnavska, Hanyun Zhang, Priya~Lakshmi Narayanan, Justin Law, Yinyin Yuan, Abhiroop Tejomay, Aditya Mitkari, Dinesh Koka, Vikas Ramachandra, Lata Kini, and Amit Sethi.
\newblock Monusac2020: A multi-organ nuclei segmentation and classification challenge.
\newblock \emph{IEEE Transactions on Medical Imaging}, 40\penalty0 (12):\penalty0 3413–3423, December 2021.
\newblock ISSN 1558-254X.
\newblock \doi{10.1109/tmi.2021.3085712}.
\newblock URL \url{http://doi.org/10.1109/TMI.2021.3085712}.

\bibitem[Vu et~al.(2019)Vu, Graham, Kurc, To, Shaban, Qaiser, Koohbanani, Khurram, Kalpathy-Cramer, Zhao, Gupta, Kwak, Rajpoot, Saltz, and Farahani]{cpm}
Quoc~Dang Vu, Simon Graham, Tahsin Kurc, Minh Nguyen~Nhat To, Muhammad Shaban, Talha Qaiser, Navid~Alemi Koohbanani, Syed~Ali Khurram, Jayashree Kalpathy-Cramer, Tianhao Zhao, Rajarsi Gupta, Jin~Tae Kwak, Nasir Rajpoot, Joel Saltz, and Keyvan Farahani.
\newblock Methods for segmentation and classification of digital microscopy tissue images.
\newblock \emph{Frontiers in Bioengineering and Biotechnology}, 7, April 2019.
\newblock ISSN 2296-4185.
\newblock \doi{10.3389/fbioe.2019.00053}.
\newblock URL \url{http://doi.org/10.3389/fbioe.2019.00053}.

\bibitem[Wang et~al.(2024)Wang, Qiu, Hao, Jin, Gao, Qi, Xu, Zhang, and Xu]{srsa-net}
Ranran Wang, Yusong Qiu, Xinyu Hao, Shan Jin, Junxiu Gao, Heng Qi, Qi~Xu, Yong Zhang, and Hongming Xu.
\newblock Simultaneously segmenting and classifying cell nuclei by using multi-task learning in multiplex immunohistochemical tissue microarray sections.
\newblock \emph{Biomedical Signal Processing and Control}, 93:\penalty0 106143, July 2024.
\newblock ISSN 1746-8094.
\newblock \doi{10.1016/j.bspc.2024.106143}.
\newblock URL \url{http://doi.org/10.1016/j.bspc.2024.106143}.

\bibitem[Weigert and Schmidt(2022)]{stardist-histopatho}
Martin Weigert and Uwe Schmidt.
\newblock Nuclei instance segmentation and classification in histopathology images with stardist.
\newblock In \emph{2022 IEEE International Symposium on Biomedical Imaging Challenges (ISBIC)}, page 1–4. IEEE, March 2022.
\newblock \doi{10.1109/isbic56247.2022.9854534}.
\newblock URL \url{http://doi.org/10.1109/ISBIC56247.2022.9854534}.

\bibitem[Xu et~al.(2024)Xu, Goetz, and Rajpoot]{nasir-sam}
Kesi Xu, Lea Goetz, and Nasir Rajpoot.
\newblock On generalisability of segment anything model for nuclear instance segmentation in histology images, 2024.
\newblock URL \url{https://arxiv.org/abs/2401.14248}.

\bibitem[Yue et~al.(2024)Yue, Zhang, Hu, Xia, Luo, and Wang]{surgical-sam}
Wenxi Yue, Jing Zhang, Kun Hu, Yong Xia, Jiebo Luo, and Zhiyong Wang.
\newblock Surgicalsam: Efficient class promptable surgical instrument segmentation.
\newblock \emph{Proceedings of the AAAI Conference on Artificial Intelligence}, 38\penalty0 (7):\penalty0 6890–6898, March 2024.
\newblock ISSN 2159-5399.
\newblock \doi{10.1609/aaai.v38i7.28514}.
\newblock URL \url{http://doi.org/10.1609/aaai.v38i7.28514}.

\bibitem[Zhang et~al.(2023)Zhang, Ma, Kapse, Saltz, Vakalopoulou, Prasanna, and Samaras]{sam-path}
Jingwei Zhang, Ke~Ma, Saarthak Kapse, Joel Saltz, Maria Vakalopoulou, Prateek Prasanna, and Dimitris Samaras.
\newblock \emph{SAM-Path: A Segment Anything Model for Semantic Segmentation in Digital Pathology}, page 161–170.
\newblock Springer Nature Switzerland, 2023.
\newblock ISBN 9783031474019.
\newblock \doi{10.1007/978-3-031-47401-9_16}.
\newblock URL \url{http://doi.org/10.1007/978-3-031-47401-9_16}.

\bibitem[Zhang et~al.(2024)Zhang, Li, Xue, Wang, and Li]{zhang2024glandsam}
Qixiang Zhang, Yi~Li, Cheng Xue, Haonan Wang, and Xiaomeng Li.
\newblock Glandsam: Injecting morphology knowledge into segment anything model for label-free gland segmentation.
\newblock \emph{IEEE Transactions on Medical Imaging}, 2024.
\newblock URL \url{https://doi.org/10.1109/TMI.2024.3476176}.

\bibitem[Zhao et~al.(2024)Zhao, Gu, Yang, Usuyama, Lee, Kiblawi, Naumann, Gao, Crabtree, Abel, Moung-Wen, Piening, Bifulco, Wei, Poon, and Wang]{biomedparse}
Theodore Zhao, Yu~Gu, Jianwei Yang, Naoto Usuyama, Ho~Hin Lee, Sid Kiblawi, Tristan Naumann, Jianfeng Gao, Angela Crabtree, Jacob Abel, Christine Moung-Wen, Brian Piening, Carlo Bifulco, Mu~Wei, Hoifung Poon, and Sheng Wang.
\newblock A foundation model for joint segmentation, detection and recognition of biomedical objects across nine modalities.
\newblock \emph{Nature Methods}, November 2024.
\newblock ISSN 1548-7105.
\newblock \doi{10.1038/s41592-024-02499-w}.
\newblock URL \url{http://doi.org/10.1038/s41592-024-02499-w}.

\bibitem[Zou et~al.(2024)Zou, Yang, Zhang, Li, Li, Wang, Wang, Gao, and Lee]{seem}
Xueyan Zou, Jianwei Yang, Hao Zhang, Feng Li, Linjie Li, Jianfeng Wang, Lijuan Wang, Jianfeng Gao, and Yong~Jae Lee.
\newblock Segment everything everywhere all at once.
\newblock In \emph{Proceedings of the 37th International Conference on Neural Information Processing Systems}, NIPS '23, Red Hook, NY, USA, 2024. Curran Associates Inc.
\newblock URL \url{https://openreview.net/forum?id=UHBrWeFWlL}.

\end{thebibliography}

\appendix
\section{Evaluation Metric} \label{app:metric}

We use the mean segmentation accuracy, following \cite{dsb}, to evaluate instance segmentation results.
It is based on true positives ($TP$), false negatives ($FN$), and false positives ($FP$), which are derived from the intersection over union (IoU) of predicted and true objects.
Specifically, $TP(t)$ is defined as the number of matches between predicted and true objects with an IoU above the threshold $t$, $FP(t)$ correspond to the number of predicted objects minus $TP(t)$, and $FN(t)$ to the number of true objects minus $TP(t)$.
The mean segmentation accuracy is computed over multiple thresholds:
\begin{equation*}
    \text{Mean Segmentation Accuracy} = \frac{1}{|\text{\# thresholds}|} \sum_{t} \frac{TP(t)}{TP(t) + FP(t) + FN(t)}\,.
\end{equation*}
Here, we use thresholds $t \in [0.5, 0.55, 0.6, 0.65, 0.7, 0.75, 0.8, 0.85, 0.9, 0.95]$. For each dataset, we report the average mean segmentation accuracy over images in the test set.

We use the dice coefficient to evaluate the semantic segmentation results. It is defined as:
\begin{equation*}
    \text{Dice Coefficient} = \frac{2 * \sum{p_i \, t_i}}{\sum p_i + \sum t_i},
\end{equation*}
for per-pixel prediction values $p_i$ and per pixel target values $t_i$. When reporting overall results for semantic segmentation, we compute the dice coefficient per class and then take a weighted sum of coefficients, with the weight corresponding to the class frequency.

Note that prior work regularly reports the panoptic quality (PQ) as metric for semantic instance segmentation on PanNuke, the dataset we use to evaluate the semantic segmentation.
However, we reviewed the referenced code at \url{https://github.com/TissueImageAnalytics/PanNuke-metrics/blob/master/utils.py} and found that it is not a proper implementation of the PQ, and that it seems to only evaluate the semantic segmentation.
Hence, we decided to use the Dice coefficient, which is a more established metric for semantic segmentation.
We will investigate this further in future work, and if possible will introduce a true PQ metric.

\section{Dataset Description} \label{app:data}

For evaluating PathoSAM, we collected a total of 16 datasets, detailed in \ref{tab:datasets}. It comprises of datasets with different tissue samples, magnifications, stainings and dimensions. We reserve the Janowczyk dataset \cite{jano-user-study} for the user study, details provided in App. \ref{app:user-study}. Besides that, we use all datasets for either training our generalist models, evaluating them or for further downstream tasks.
 
\begin{table}[h]
\centering
\renewcommand{\arraystretch}{1.2}
\begin{tabular}{|l|l|p{6cm}|}
\hline
\textbf{Dataset} & \textbf{Tissue (Staining)} & \textbf{Num. Samples \textbar{} Magnification \newline (Avg. Dimension)} \\   
\hline
\textbf{CoNSeP} & Colon (H\&E) & 41 \textbar{} 40x (1000, 1000, 3) \\
\hline
\textit{CPM15} & Brain (H\&E) & 15 \textbar{} 40x, 20x (584, 668, 3) \\  
\hline
\textit{CPM17} & Brain (H\&E) & 64 \textbar{} 40x, 20x (500, 500, 3) \\  
\hline
\textbf{CryoNuSeg} & 10 distinct tissues (H\&E) & 30 \textbar{} 40x (512, 512, 3) \\  
\hline
\underline{GlaS} & Colon (H\&E) & 165 \textbar{} 40x (514, 760, 3) \\  
\hline
\underline{Janowczyk} & Breast (H\&E) & 143 \textbar{} 40x (2000, 2000, 3) \\  
\hline
\textit{Lizard} & Colon (H\&E) & 238 \textbar{} 20x (934, 1055, 3) \\
\hline
\textbf{LyNSec} & Lymphoma (H\&E, IHC) & 600 \textbar{} 40x (512, 512, 3) \\  
\hline
\underline{MoNuSAC} & 4 distinct tissues (H\&E) & 310 \textbar{} 40x (559, 602, 3) \\  
\hline
\textit{MoNuSeg} & 7 distinct tissues (H\&E) & 51 \textbar{} 40x (1000, 1000, 3) \\  
\hline
\underline{NuClick} & Lymphocytes (IHC) & 871 \textbar{} 40x (256, 256, 3) \\  
\hline
\textbf{NuInsSeg} & 31 distinct tissues (H\&E) & 655 \textbar{} 40x (512, 512, 3) \\  
\hline
\textit{PanNuke} & 19 distinct tissues (H\&E) & 7801 \textbar{} 40x (256, 256, 3) \\  
\hline
\textit{PUMA} & Melanoma (H\&E) & 206 \textbar{} 40x (1024, 1024, 3) \\  
\hline
\textbf{IHC TMA} & Lung (IHC) & 266 \textbar{} 40x (256, 256, 3) \\  
\hline
\textbf{TNBC} & Breast (H\&E) & 50 \textbar{} 40x (512, 512, 3) \\  
\hline
\end{tabular}
\caption{Description of the different datasets used in our study for training generalist models (marked in \textit{italics}), evaluating (marked in \textbf{bold}) and comparing them to other methods, and evaluating downstream tasks (marked in \underline{underline}).}
\label{tab:datasets}
\end{table}

\section{Architecture and Training Details} \label{app:arch_train}

\begin{figure}[htbp]
\label{fig:architecture}
\floatconts
  {fig:architecture}
  {\caption{a) The architecture of our generalist model, which combines SAM with an additional decoder for instance segmentation, following the design from $\mu$SAM. b) The architecture for semantic segmentation, which uses a ViT-B image encoder, initialized with the weights of the PathoSAM generalist model, and an additional convolutional decoder that predicts semantic class probabilities.}}
  {\includegraphics[width=\linewidth]{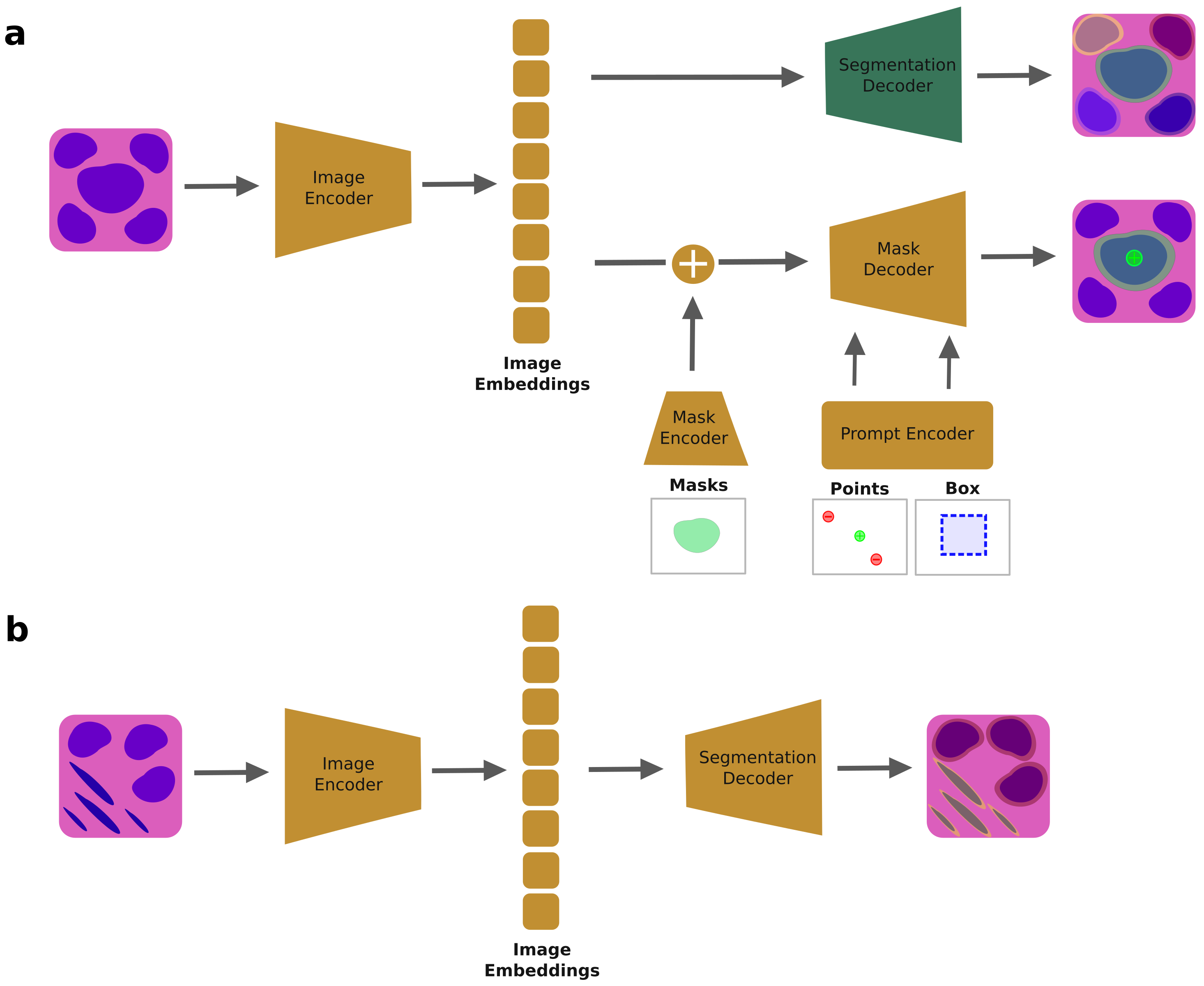}}
\end{figure}

The architecture of our generalist models is shown in Fig.~\ref{fig:architecture} a).
It uses the original SAM architecture with the additional instance segmentation decoder, following the design of $\mu$SAM,
see also Sec.~\ref{sec:method_ais}.

The architecture for semantic segmentation based on the generalist is shown in Fig.~\ref{fig:architecture} b).
For our experiments, we use the image encoder initialized with pretrained PathoSAM model weights; the additional convolutional decoder predicts semantic class probabilities in independent channels.

We train the generalist model with the objective from $\mu$SAM \cite{micro-sam} that combines training of interactive and automatic segmentation (see Sec.~\ref{sec:method_sam} and Sec.~\ref{sec:method_ais}).
The generalist models are trained for 250,000 iterations, the PanNuke model and the specialist models for downstream tasks are trained for 100,00 iterations.
We use 7 correction iterations in the training objective and a probability for using a mask prompt of 50\% per iteration.
This choice enables interactive both with and without addition of the mask prompt. 
We use a batch size of 2 and train with 40 (ViT-B), 30 (ViT-L), or 25 (ViT-H) objects per image in the batch as hyperparameters for the training objective.
We use the AdamW optimizer with an initial learning rate of 1e-5. We use a learning rate scheduler that scales it by a factor of 0.9 when the validation metric plateuas for 10 epochs. All models are trained on a single NVIDIA A100 GPU with 80GB VRAM.

\section{Segmentation Quality}

\subsection{Automatic Segmentation} \label{app:autoseg}

\begin{figure}[htbp]
\label{fig:autoseg-quali}
\floatconts
  {fig:example}
  {\caption{Qualitative plots for automatic instance segmentation with PathoSAM and other methods. Names of datasets in italics are in-domain, and bold are out-of-domain.}}
  {\includegraphics[width=0.75\linewidth]{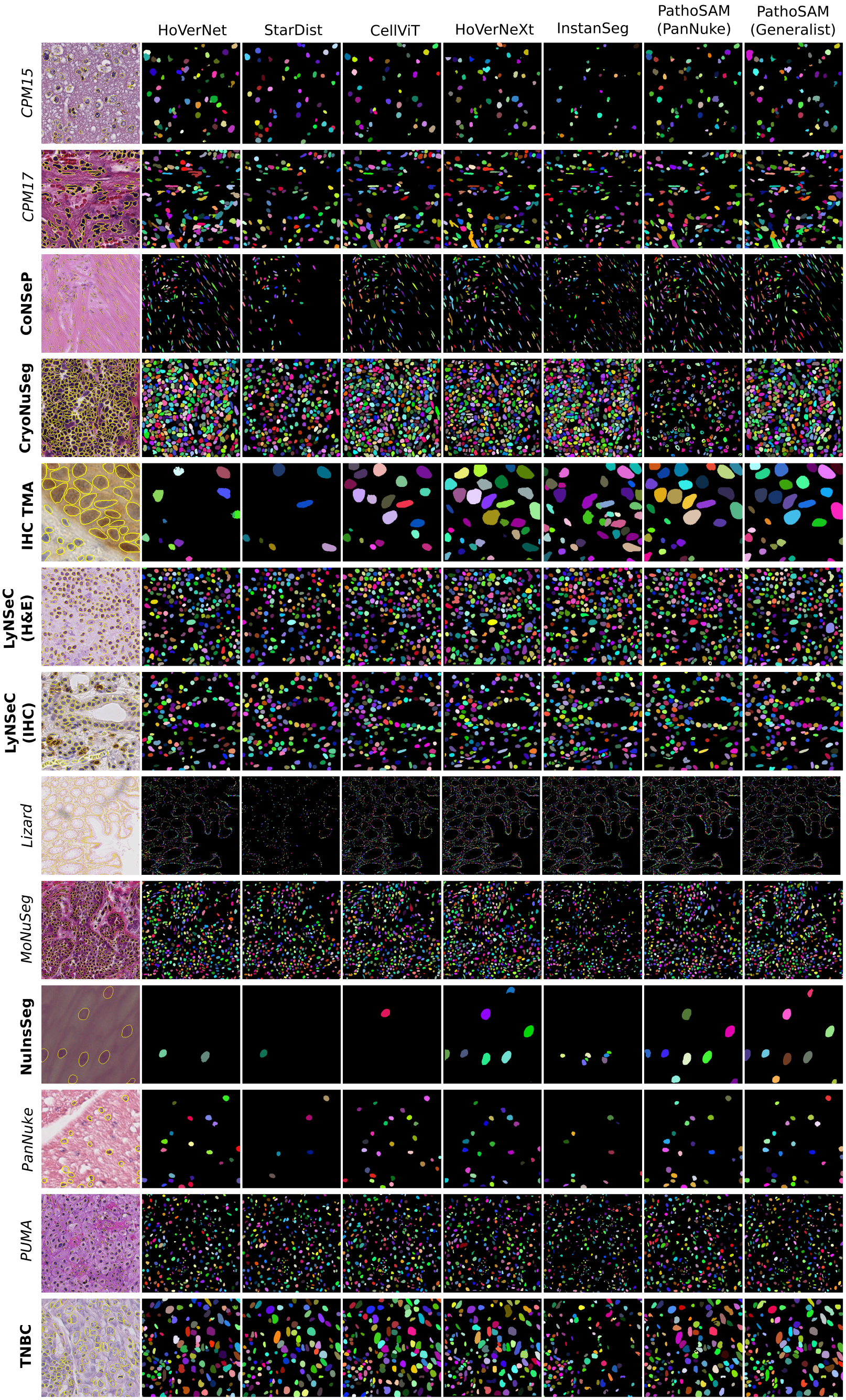}}
\end{figure}

\begin{figure}[htbp]
\label{fig:autoseg-quanti}
\floatconts
  {fig:example}
  {\caption{Quantitative results for automatic instance segmentation with PathoSAM and all other methods. Names of datasets in italics are in-domain, and bold are out-of-domain.}}
  {\includegraphics[width=\linewidth]{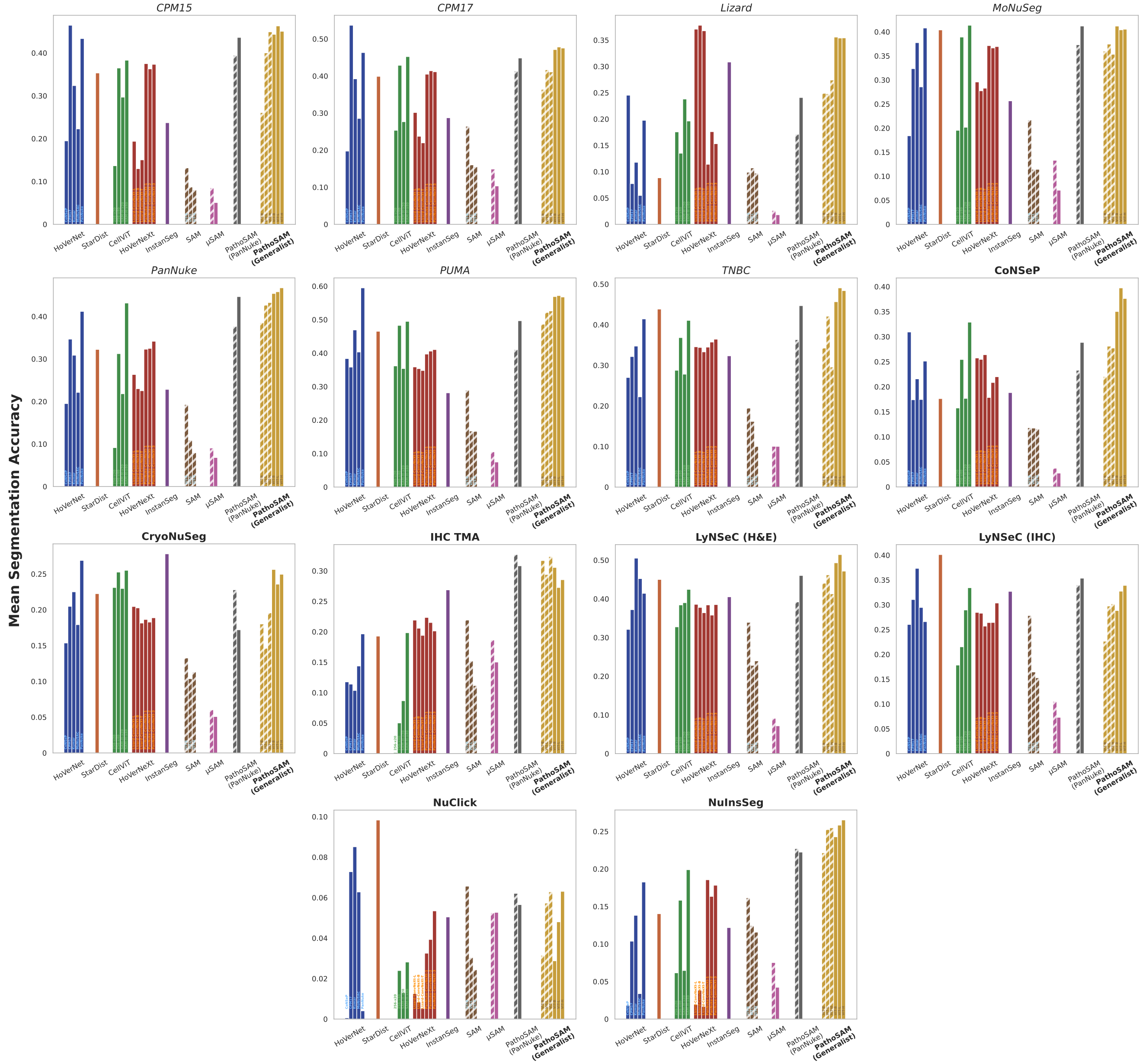}}
\end{figure}

We provide comprehensive overview of all results for automatic instance segmentation. \ref{fig:autoseg-quali} displays results for all 8 datasets discussed in the main paper, and additional 5 datasets, which include CPM15 \cite{cpm}, CPM17 \cite{cpm}, CoNSeP \cite{hovernet}, LyNSeC (IHC) \cite{lynsec} and TNBC \cite{tnbc}.
Further, \ref{fig:autoseg-quanti} includes detailed quantitative results for all the datasets. We provide results for all combination of models for the benchmarked methods, including all trained PathoSAM generalist models. We observe that PathoSAM is consistently either the best or among the best models for the datasets, including for out-of-domain datasets that contain different stainings (e.g. IHC staining) or imaging conditions (e.g. CryoNuSeg).
For CryoNuSeg, which contains images of tissues that were cryo-sectioned, the automatic segmentation quality of the generalist model suffers from this domain shift. They also have low scores for the automatic segmentation of lymphocytes in NuClick,
which is due to the fact that other nuclei are segmented as well.
This dataset is used for further investigation in Sec. \ref{sec:res_downstream} to understand if training a tailored specialist model leads to an improved automatic lymphocytes segmentation.

\begin{figure}[htbp]
\label{fig:cryonuseg-samples}
\floatconts
  {fig:example}
  {\caption{Three example of CryoNuSeg images for with ground-truth nucleus annotations and automatic segmentation from PathoSAM. The white box highlight the following challenges in this dataset: a) PathoSAM scores bad where the annotations are inconsistent with the nucleus ROI, b) PathoSAM struggles in regions where the foreground is hardly visible in nuclei clusters, c) PathoSAM performs well for most objects, but the scores drop due to overlapping nucleus annotations.}}
  {\includegraphics[width=\linewidth]{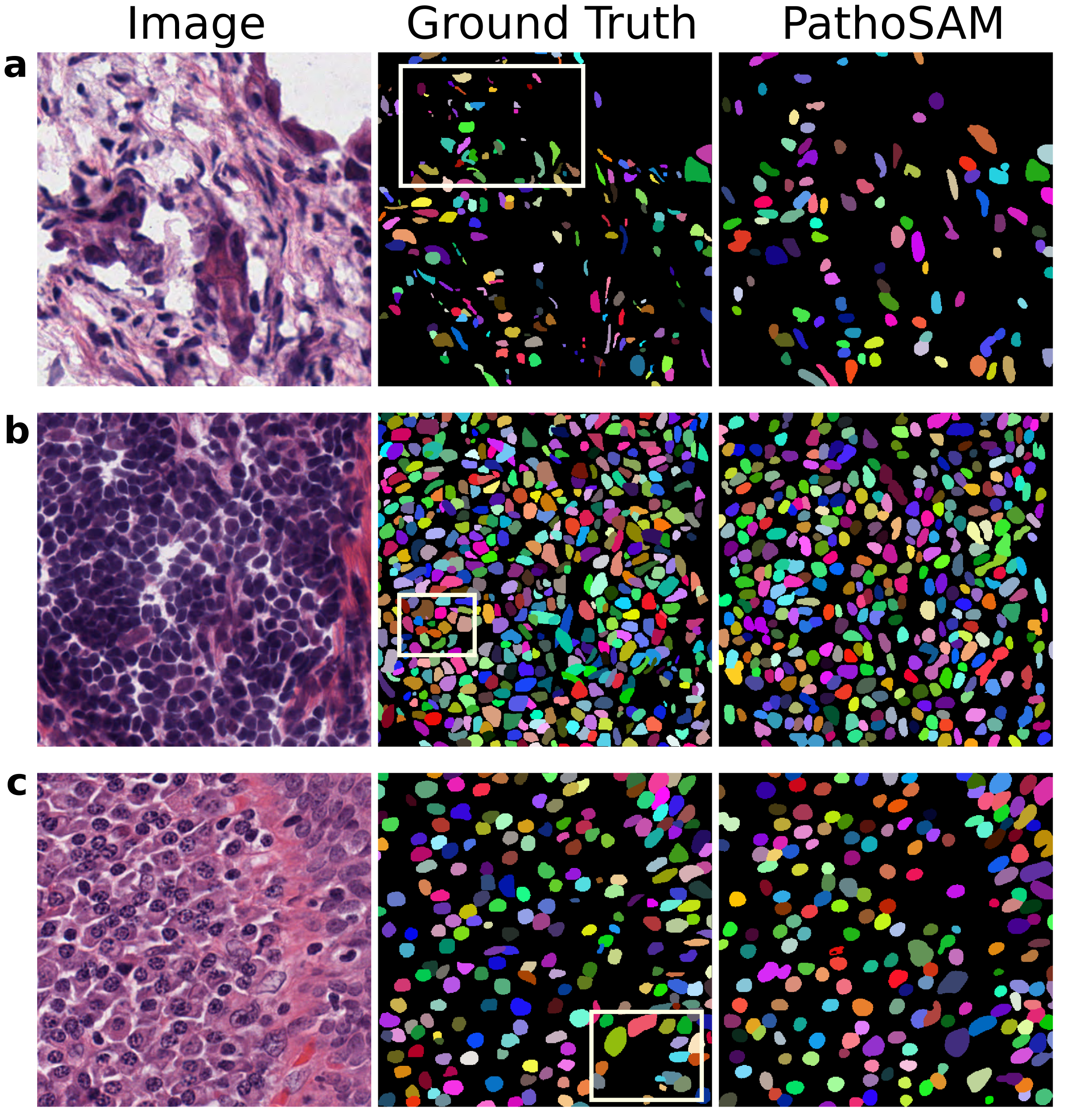}}
\end{figure}
 
\begin{figure}[htbp]
\label{fig:autoseg-f1-precision-recall}
\floatconts
  {fig:autoseg-f1-precision-recall}
  {\caption{F1 Score, Recall and Precision for IoU thresholds from 0.5 to 0.95, for all datasets, for automatic instance segmentation with PathoSAM model (Generalist - ViT-L). Names of datasets in italics are in-domain, and bold are out-of-domain.}}
  {\includegraphics[width=\linewidth]{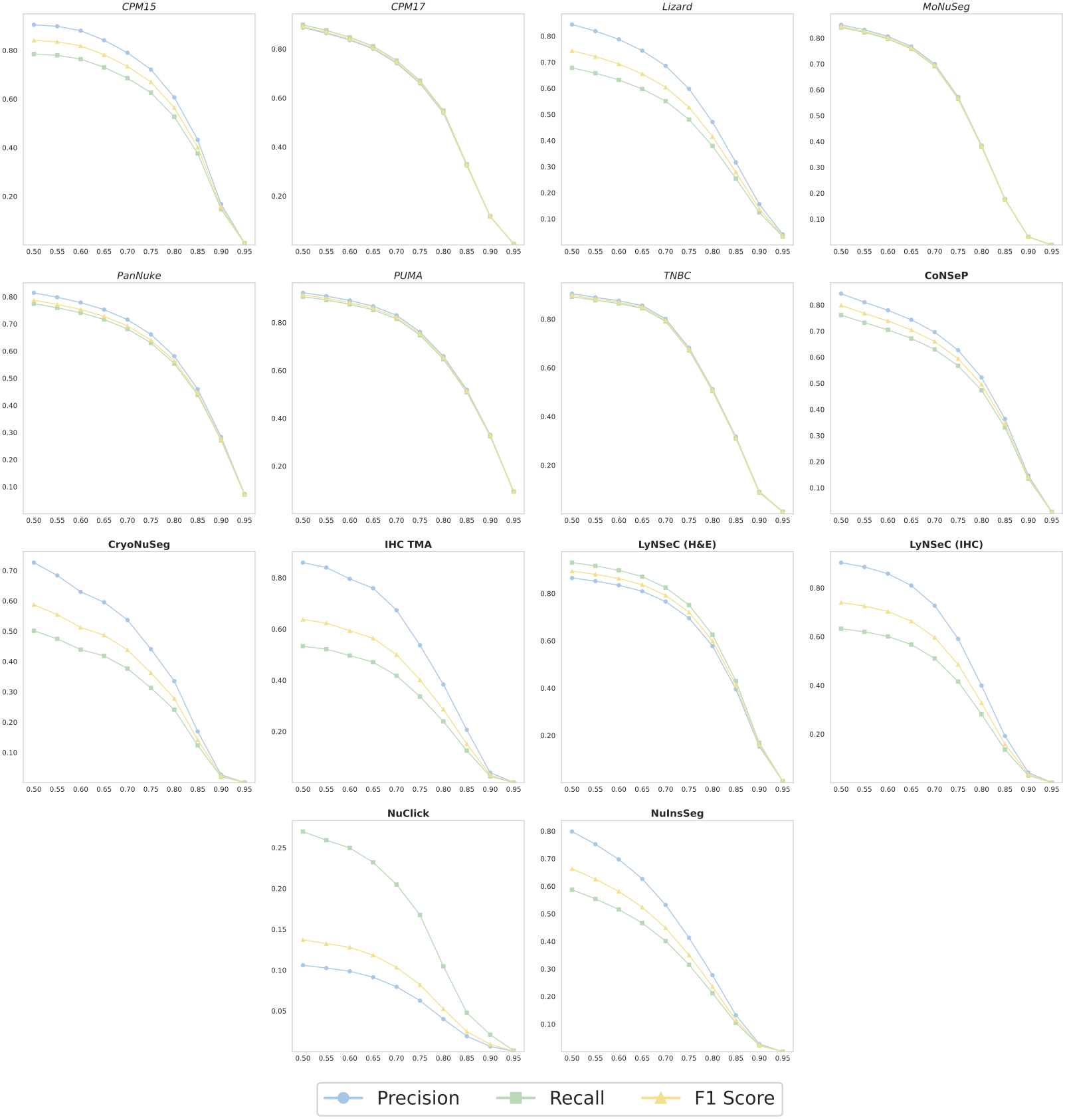}}
\end{figure}

\begin{figure}[htbp]
\label{fig:heatmap-lynsec-he}
\floatconts
  {fig:heatmap-lynsec-he}
  {\caption{Grid search results over (the most important) parameters of the watershed used to obtain the instance segmentation in AIS. These parameters are the thresholds applied to center and boundary distances predicted by the segmentation decoder in order to find seeds. Please refer to \cite{micro-sam} for details on the instance segmentation methodology. We report the F1 Score, recall and precision for a grid-search over the two thresholds on 15 validation images of the LyNSeC (H\&E) dataset, i.e. an OOD dataset. We see that precision is best for small (boundary) threshold values whereas recall is best for large thresholds. This is due to the fact that large threshold values lead to multiple seeds being derived from the distance maps, resulting in over-segmentation. Low threshold values have the opposite effect and result in under-segmentation. Hence, tuning the (boundary) threshold enables varying the degree of over-/under-segmentation.}}
  {\includegraphics[width=\linewidth]{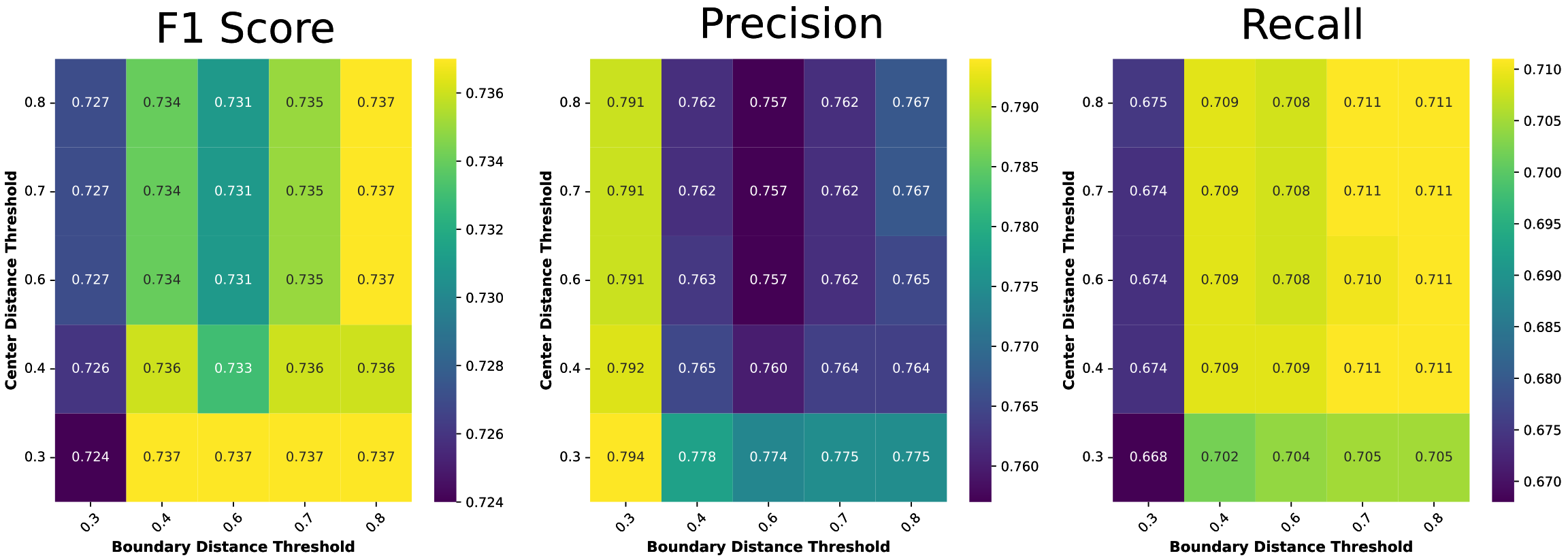}}
\end{figure}

\begin{figure}[htbp]
\label{fig:neutrophils-seg}
\floatconts
  {fig:neutrophils-seg}
  {\caption{Qualitative and quantitative results of PathoSAM on neutrophil granulocytes, and a further specialist model trained on a few images for neutrophil granulocyte segmentation on \cite{puma}. Neutrophils have multi-lobed nuclei, which are more challenging to segment compared to ``regular'' nuclei. a) The scores are on par, with a slight favor for the specialist model. Note: we report scores specifically for neutrophil granulocyte instance segmentation, excluding other nucleus types from the evaluation. b) An example image with annotations for neutrophils. The zoom-ins show the segmentation results for the generalist model (middle) and the neutrophil specialist model (right). Arrows indicate neutrophils that are segmented better by the specialist compared to the generalist. These results indicate that both models can segment neutrophils reasonably well, and that the specialist model performs a bit better than our generalist. Importantly, this shows that our method can segment such complex shapes correctly, and that in practice the quality of results depends on the presence and quality of the respective annotations in the training data.}}
  {\includegraphics[width=\linewidth]{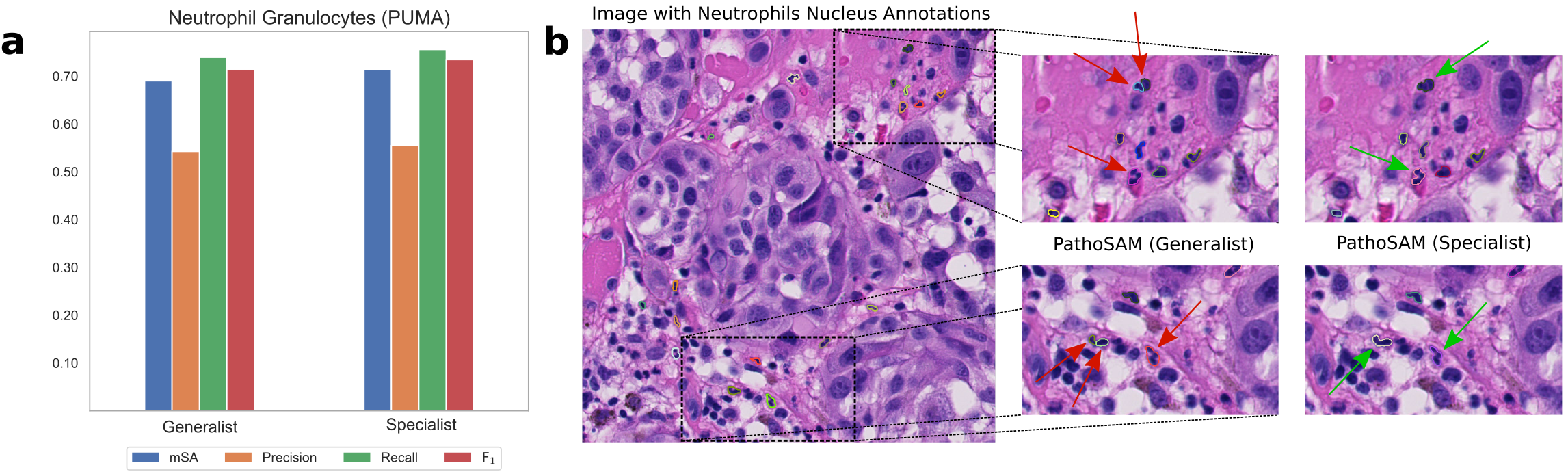}}
\end{figure}

\begin{figure}[htbp]
\label{fig:semantic-seg}
\floatconts
  {fig:semantic_seg}
  {\caption{Quantitative results for semantic segmentation, for different versions of the benchmarked methods (a) and the per-class performance for the respective best method (b). c) shows qualitative examples.}}
  {\includegraphics[width=\linewidth]{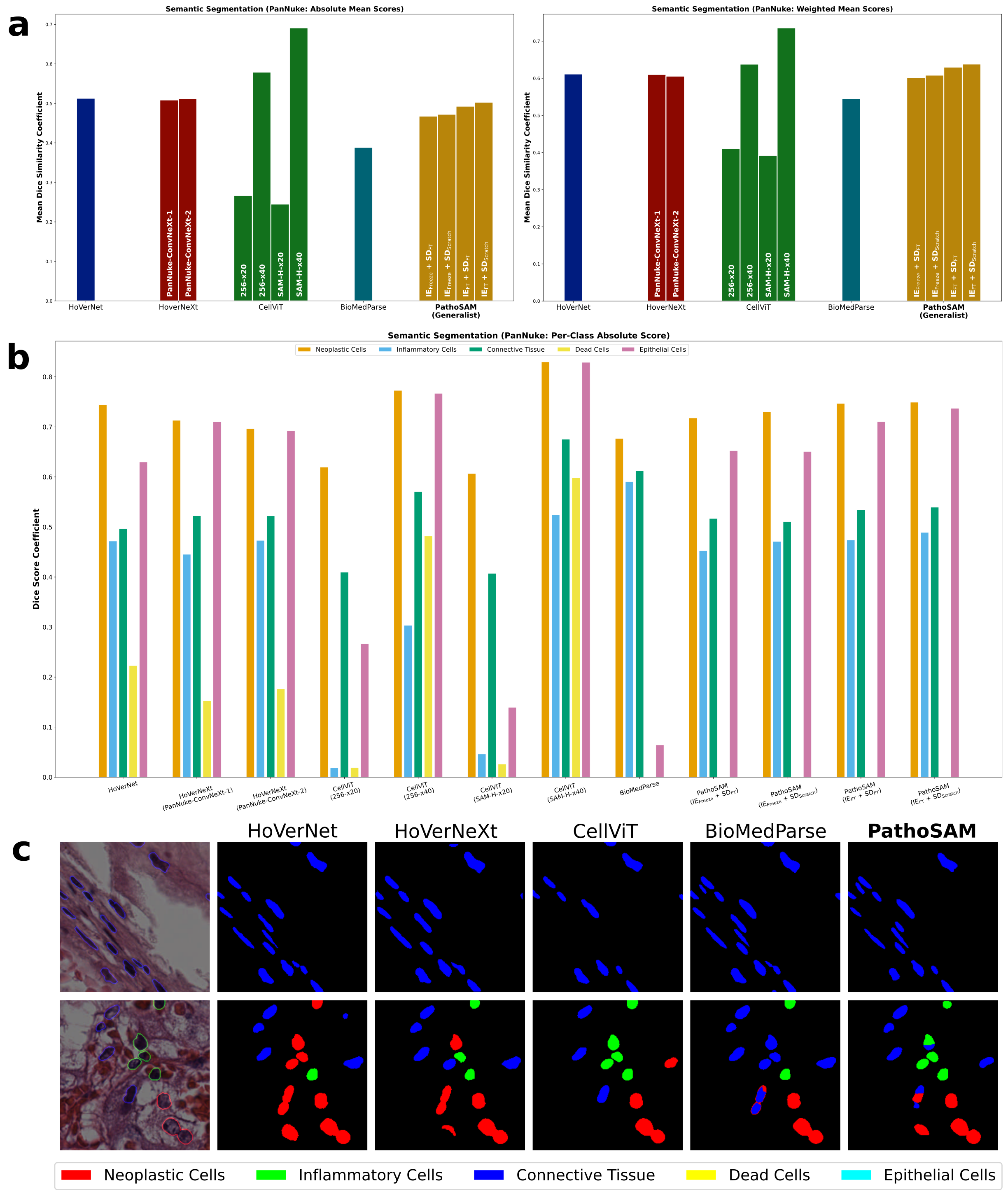}}
  \label{fig:semantic_seg}
\end{figure}

For semantic segmentation we compare different methods that were trained for this task on the PanNuke dataset with our models finetuned based on the PathoSAM generalist (ViT-B).
The results are described and discussed in Sec.~\ref{sec:res_downstream}. 
For BioMedParse, the model iteratively auto-prompts itself with modality-specific \textit{predefined} text prompts. In this case, the model only requires the input image and the corresponding modality name (i.e. \textit{Pathology} for our experiments). It performs an additional post-processing step to resolve potential overlaps, and combines all generated masks.
Fig.~\ref{fig:semantic_seg} shows a more detailed qualitative and quantitative overview of the results.

\subsection{Interactive Segmentation} \label{app:intseg}

\begin{figure}[htbp]
\label{fig:intseg-quali}
\floatconts
  {fig:intseg-quali}
  {\caption{Qualitative plots for interactive segmentation with SAM, $\mu$SAM and PathoSAM ViT-B generalist model.}}
  {\includegraphics[width=\linewidth]{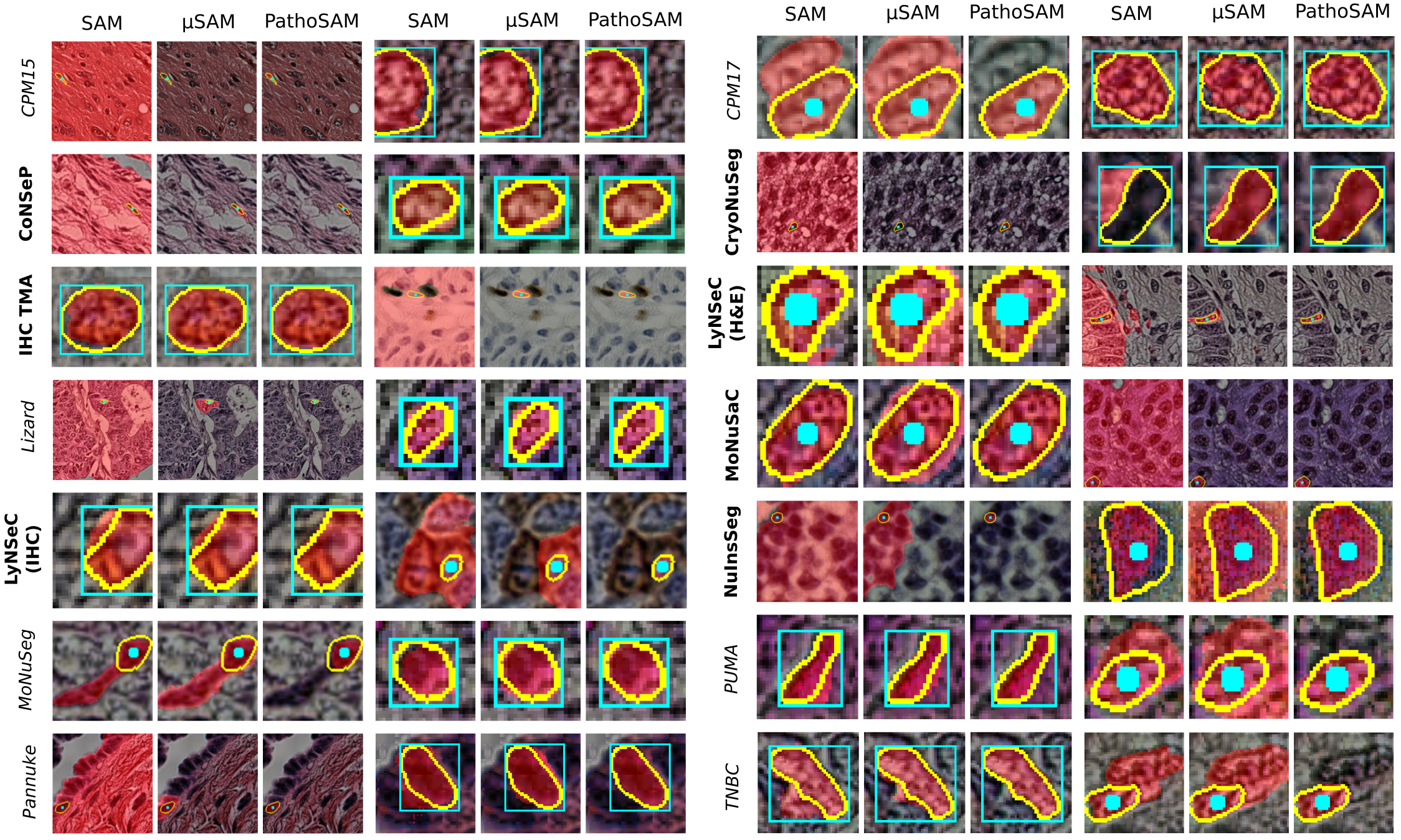}}
\end{figure}

We provide a comprehensive overview of all results for interactive segmentation. Fig.~\ref{fig:intseg-quali} displays results for all 8 datasets discussed in the main paper, and 6 additional datasets, including CPM15 \cite{cpm}, CPM17 \cite{cpm}, CoNSeP \cite{hovernet}, MoNuSaC \cite{monusac}, LyNSeC (IHC) \cite{lynsec} and TNBC \cite{tnbc}. Notably, MoNuSaC has additional ambiguous labels, including annotations for the cells or cytoplasm. However, we decided to include it in our evaluation to evaluate it for interactive segmentation.
We clearly observe that the PathoSAM generalist outperforms SAM and $\mu$SAM for interactive segmentation with a single point or box. Considering that the model has not been trained on any other staining besides H\&E, the results for IHC stained images are convincing.

\begin{figure}[htbp]
\label{fig:intseg-quanti}
\floatconts
  {fig:example}
  {\caption{Quantitative results for interactive segmentation for SAM, $\mu$SAM, PathoSAM (PanNuke) and all variants of PathoSAM (Generalist) models.}}
  {\includegraphics[width=\linewidth]{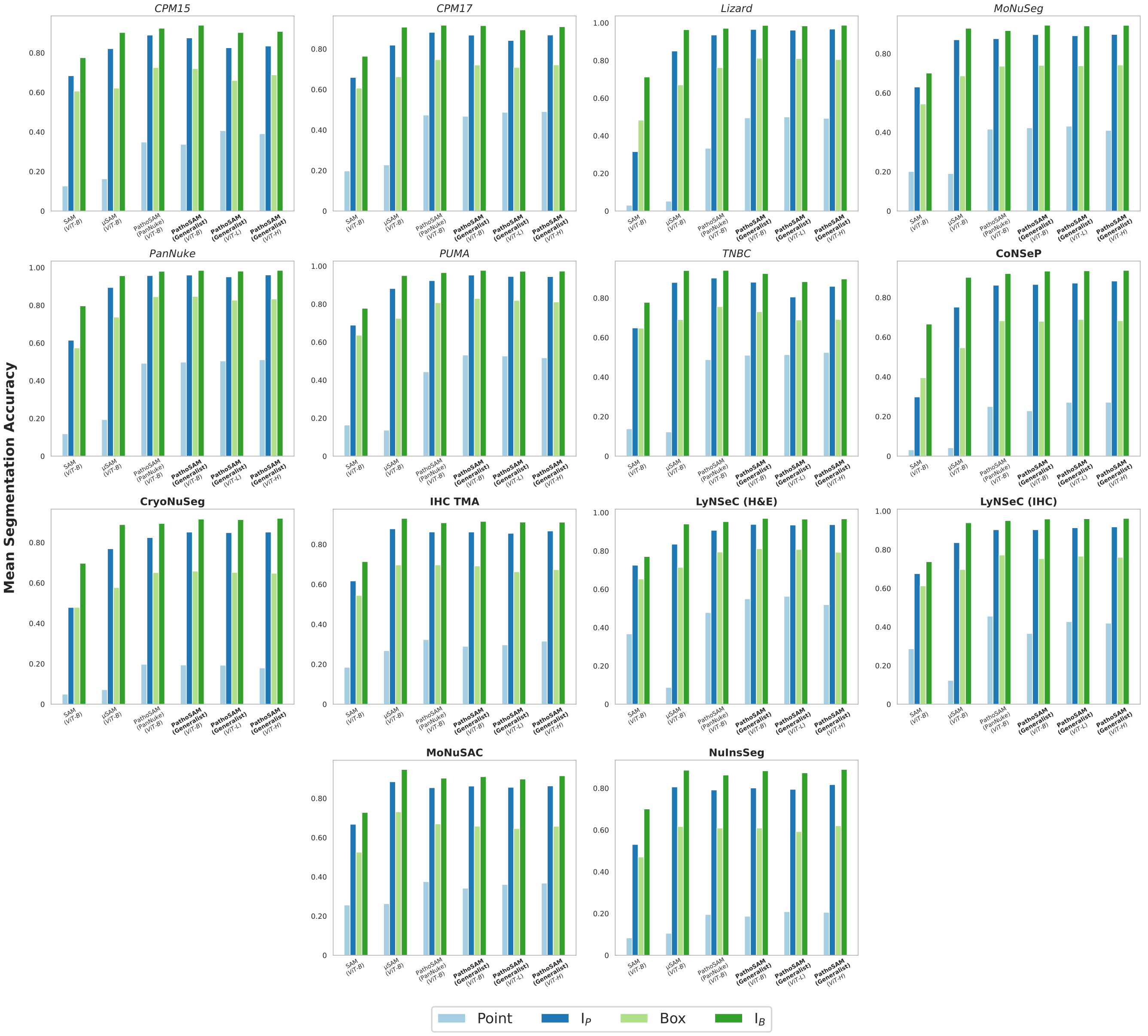}}
\end{figure}

Further, Fig.~\ref{fig:intseg-quanti} shows detailed quantitative results for all the above mentioned datasets. We provide quantitative results for all trained PathoSAM generalist models.
Overall, we observe that our generalist model clearly outperforms SAM and $\mu$SAM, and the segmentation quality is consistent along different variants of the generalist model.
For some datasets, the model trained on PanNuke \cite{pannuke} performs a bit better for iterative refinement, a likely reason could be the models' distribution proximity to the training data.

\subsection{User Study} \label{app:user-study}

\begin{figure}[htbp]
\label{fig:user_study_qupath}
\floatconts
  {fig:user_study_qupath}
  {\caption{User study with $\mu$SAM in napari, with tiling-window based inference for this large image}}
  {\includegraphics[width=\linewidth]{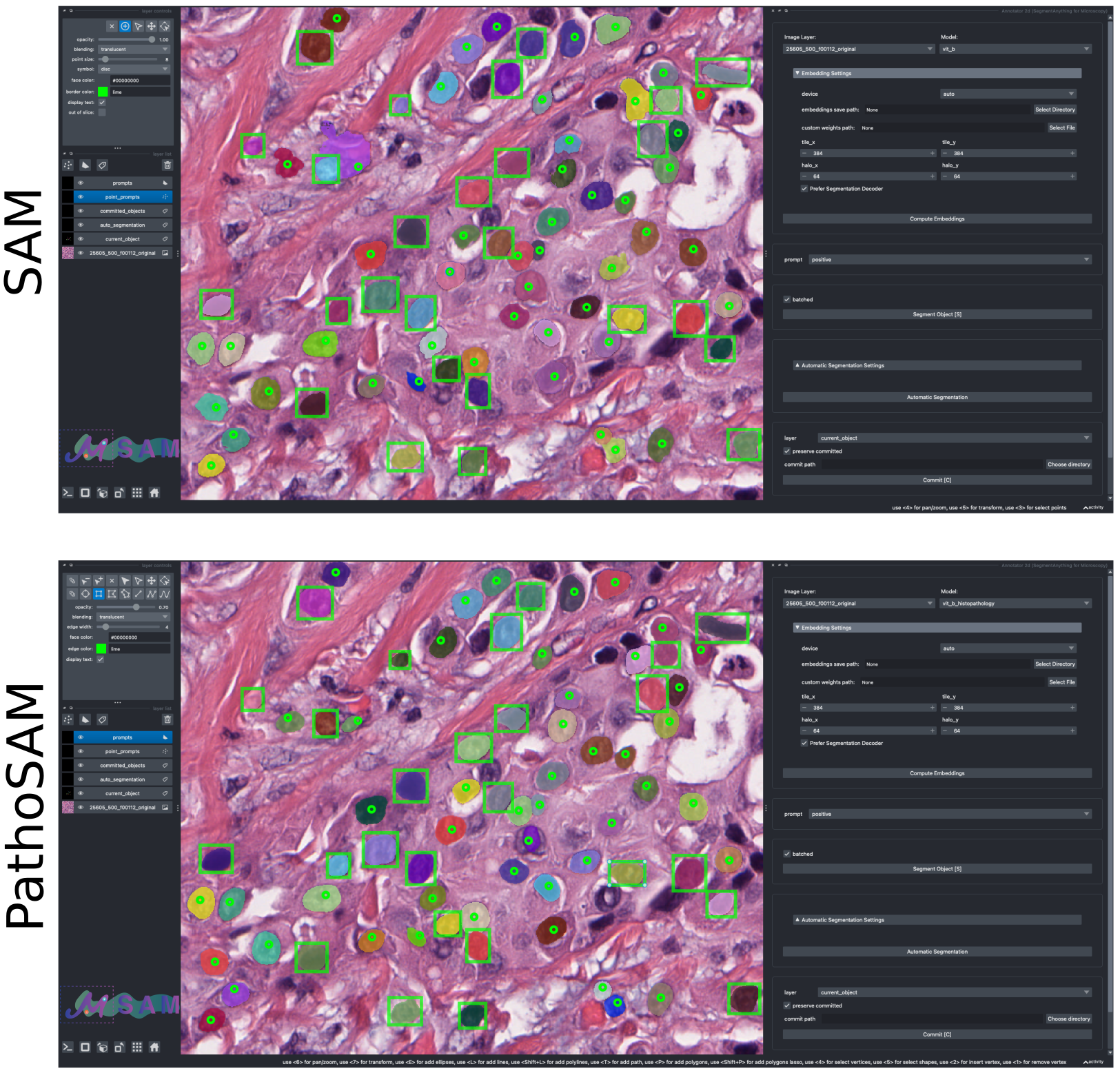}}
\end{figure}

Here, we choose a H\&E stained histopathology tissue image for validating SAM and PathoSAM for user-based interactive segmentation in two different tools.
QuPath \cite{qupath} expects the installation of SAM-API, which is well documented at \url{https://github.com/ksugar/samapi}. After starting the SAM-API server, the user can access the SAM extension and begin annotating. The tool currently supports default SAM \cite{sam} and SAM2 \cite{sam2} models. If desired, users can provide the URL for remotely stored model checkpoints, which makes it compatible with SAM-family models. 
Fig.~\ref{fig:user_study_qupath} displays a screenshot on top for annotation with default SAM model, and bottom for annotation with PathoSAM model. We clearly see that the model identifies nuclei better with our PathoSAM generalist model, using a single point or box per object.
$\mu$SAM comes with the installation package for PathoSAM, also well documented at \url{https://computational-cell-analytics.github.io/micro-sam/} for stand-alone usage. A user has the flexibility to start napari \cite{napari} and select the \textit{Segment Anything for Microscopy} available under extensions and a desired annotator (for histopathology images, the relevant choice is "Annotator 2D"), or access the 2D annotator with CLI scripts. For images larger than the training data, which is quite common for WSIs, $\mu$SAM supports segmentation over tiles over the entire image. We make use of this feature to evaluate both SAM and PathoSAM models. In addition, PathoSAM provides a simple CLI script to run automatic segmentation over WSIs.

Overall, our observation affirms PathoSAM's interactive segmentation capability and tool compatibility across popular annotation tools. We believe PathoSAM will speed up such annotation workflows.

\begin{figure}[htbp]
\label{fig:user_study_micro_sam}
\floatconts
  {fig:user_study_micro_sam}
  {\caption{User study with SAM-API in QuPath}}
  {\includegraphics[width=\linewidth]{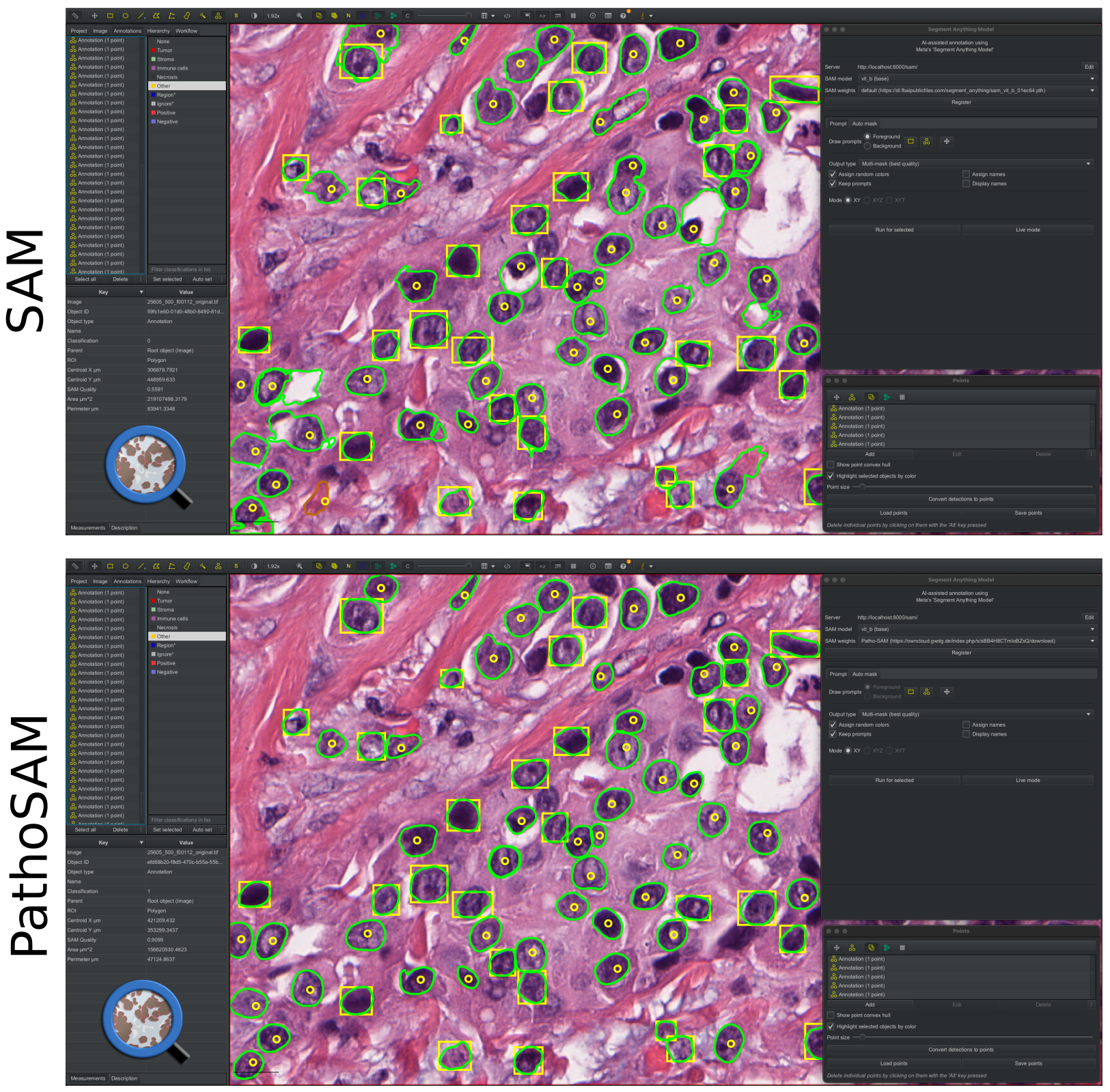}}
\end{figure}

\end{document}